\documentclass[aps,prm,reprint,runinaddress,superscriptaddress,amsmath,amssymb,floatfix,longbibliography]{revtex4-2}

\usepackage{bm}

\usepackage{color}
\usepackage{graphicx}
\usepackage{hyperref}
\usepackage{booktabs}

\usepackage{natbib}
\usepackage{braket}

\renewcommand{\vec}[1]{\boldsymbol{#1}}
\newcommand{\vhat}[1]{\vec{\hat{#1}}}

\newcommand{\subref}[2]{\ref{#1}\hyperref[#1]{#2}}

\newcommand{\byzo}{Ba$_3$Yb$_2$Zn$_5$O$_{11}$}

\newcommand{\hc}{{\rm h.c.}}
\newcommand{\muB}{\mu_{\rm B}}
\newcommand{\meV}{\ {\rm meV}}

\newcommand{\T}{\ {\rm T}}

\definecolor{cred}{RGB}{188,55,84}
\hypersetup{colorlinks=true,linkcolor=cred,citecolor=cred,urlcolor=cred}

\begin{document}

\title{Towards understanding the magnetic properties of the breathing pyrochlore compound \byzo: A single crystal study}
\author{Sachith Dissanayake}
\affiliation{Department of Physics, Duke University, Durham, NC 27708, USA}
\author{Zhenzhong Shi}
\affiliation{Department of Physics, Duke University, Durham, NC 27708, USA}
\author{Jeffrey G. Rau}
\email{jrau@uwindsor.ca}
\affiliation{Max-Planck-Institut f\"ur Physik komplexer Systeme, 01187 Dresden, Germany}
\affiliation{Department of Physics, University of Windsor, 401 Sunset Avenue, Windsor, Ontario, N9B 3P4, Canada}
\author{Rabindranath Bag}
\affiliation{Department of Physics, Duke University, Durham, NC 27708, USA}
\author{William Steinhardt}
\affiliation{Department of Physics, Duke University, Durham, NC 27708, USA}
\author{Nicholas P. Butch}
\affiliation{Center for Neutron Research, National Institute of Standards and Technology, MS 6100 Gaithersburg, Maryland 20899, USA}
\author{Matthias Frontzek }
\affiliation{Neutron Scattering Division, Oak Ridge National Laboratory, Oak Ridge, Tennessee 37831, USA}
\author{Andrey Podlesnyak }
\affiliation{Neutron Scattering Division, Oak Ridge National Laboratory, Oak Ridge, Tennessee 37831, USA}
\author{David Graf}
\affiliation{National High Magnetic Field Laboratory and Department of Physics, Florida State University, Tallahassee, Florida 32310, USA.}
\author{Casey Marjerrison}
\affiliation{Department of Physics, Duke University, Durham, NC 27708, USA}
\author{Jue Liu}
\affiliation{Neutron Scattering Division, Oak Ridge National Laboratory, Oak Ridge, Tennessee 37831, USA}
\author{Michel J.P. Gingras}
\affiliation{Department of Physics and Astronomy, University of Waterloo, Ontario, N2L 3G1, Canada}
\affiliation{Canadian Institute for Advanced Research, MaRS Centre, West Tower 661 University Ave., Suite 505, Toronto, ON, M5G 1M1, Canada}
\author{Sara Haravifard}
\email{sara.haravifard@duke.edu}
\affiliation{Department of Physics, Duke University, Durham, NC 27708, USA}
\affiliation{Department of Materials Sciences and Mechanical Engineering, Duke University, Durham, NC 27708, USA}

\date{\today}

\begin{abstract}
    \byzo{} is unique among breathing pyrochlore compounds for being in the nearly-decoupled limit where inter-tetrahedron interactions are weak, hosting isolated clusters or ``molecular magnet'' like  tetrahedra of magnetic ytterbium (Yb$^{3+}$) ions. In this work, we present the first study carried out on single-crystal samples of the breathing pyrochlore \byzo{}, using a variety of magnetometry and neutron scattering techniques along with theoretical modeling.  We employ inelastic neutron scattering to investigate the magnetic dynamics as a function of applied field (with respect to both magnitude and direction) down to a temperature of 70 mK, where inelastic scattering reveals dispersionless bands of excitations as found in earlier powder sample studies, in good agreement with a single-tetrahedron model. However, diffuse neutron scattering at zero field and dc-susceptibility at finite field exhibit features suggesting the presence of excitations at low-energy that are not captured by the single tetrahedron model. Analysis of the local structure down to 2 K via pair distribution function analysis finds no evidence of structural disorder. We conclude that effects beyond the single tetrahedron model are important in describing the low-energy, low-temperature physics of \byzo{}, but their nature remains undetermined.
\end{abstract}
\maketitle
\section{INTRODUCTION}

Frustrated magnetism is a fruitful frontier in the contemporary field of quantum materials wherein mutually incompatible interactions, driven by the arrangement of the magnetic ions and their exchange couplings, can lead to a variety of unusual phenomena~\cite{Balents2010,MilaMendelsLacroix}. Due to the dependence of the frustrated interactions on lattice geometry, materials with new lattice structures may offer untapped opportunities for the exploration of new and exotic physics. 

A long-serving example of the rich physics afforded by frustrated systems is illustrated by the A$_2^{3+}$B$_2^{4+}$O$_7$ magnetic pyrochlore oxides. In these compounds,  A is a rare-earth ion (or Y$^{3+}$) and B is typically a transition metal, with both ions potentially carrying a magnetic moment and residing on the vertices of two distinct but interpenetrating  lattices of corner-sharing tetrahedra)~\cite{GardnerRMP2010,HallasXYAnnuRevCMP,RauGingrasAnnuRevCMP}. For example, these materials have attracted interest for hosting classical spin ice physics~\cite{Bramwell1495,Springer-spin-ice} in  Ho$_2$Ti$_2$O$_7$~\cite{Harris1997,Bramwell1495,PhysRevLett.87.047205} and Dy$_2$Ti$_2$O$_7$~\cite{Ramirez1999} and potentially their quantum analogues~\cite{Gingras_2014,RauGingrasAnnuRevCMP}, unconventional ordered states~\cite{ChangHiggs2012,StewartJPCM2004,Javanparast}, order-by-disorder as in Er$_2$Ti$_2$O$_7$~\cite{Zhitomirsky2012,Savary_Er2ti2O7,PhysRevB.93.184408,Oitmaa} and possibly other, more exotic, quantum spin liquids~~\cite{PhysRevLett.82.1012,KimuraQuantum2013,GaoNatPhys2019}. 

Over the past few years, a new related class of systems, the ``breathing'' pyrochlores, has garnered excitement.  In these, the alternating corner-sharing tetrahedra of the lattice are distinguished by two different bond lengths, breaking inversion symmetry and leading to competition between inter- and intra-tetrahedron exchange interactions. The difference in size between the two types of tetrahedra can be characterized by the ``breathing ratio'', defined as $d'/d$ where $d$ and $d'$ 
are the nearest-neighbor bond distances between magnetic ions in small and large tetrahedra, respectively.
Unfortunately, studies of breathing pyrochlores have been limited to a small number of systems due the scarcity of successfully synthesized compounds, with single crystal samples being even more elusive. 

One family of breathing pyrochlore materials that has been studied in detail is the Li(In,Ga)Cr$_4$(O,S,Se)$_8$ spinels~\cite{LeePRB2016,Okamoto2013,OkamotoJPSJ2015,OkamotoJPSJ2018,PokharelPRB2016}. These compounds have breathing ratios of $d'/d \approx 1$ and are thus in a regime of effective (competing) interactions close to the regular, non-breathing, pyrochlore lattice. LiInCr$_4$S$_8$, LiGaCr$_4$S$_8$ and CuInCr$_4$S$_8$ (first characterized in the 1970s~\cite{PINCH1970425}) are few members of this family which have been synthesized successfully. While these materials have provided interesting opportunities to study inter-tetrahedron coupling effects for the case of $d'/d \approx 1$ where the ``up'' and ``down'' tetrahedra have similar sizes, they leave unanswered questions about what may be possible with a more exaggerated breathing ratio. The large $S=3/2$ spin for the Cr$^{3+}$ ions in these materials may also suppress quantum effectss~\cite{GhoshNPJQM2019}; an $S=1/2$ breathing pyrochlore magnet is less likely to be classical.
  
Study of the Li(In,Ga)Cr$_4$(O,S,Se)$_8$ has also been hindered by the lack of single crystal samples, which complicates the determination of the magnetic Hamiltonians of these systems~\cite{GhoshNPJQM2019}. \byzo{}, however, lies at the opposite extreme where the breathing ratio is large, $d'/d \approx 2$~\cite{KimuraPRB2014}. This large breathing ratio leads to very different sizes for the ``up'' and ``down'' tetrahedra, resulting in rather isolated small tetrahedra with weak inter-tetrahedron interactions. Characterization of \byzo{}~\cite{KimuraPRB2014,RauPRL2016,HakuPRB2016,HakuJPSJ2016,Park2016} has confirmed this picture of an interesting kind of ``molecular magnet'' of largely, it appears, independent tetrahedral spin clusters, each exhibiting a pair of degenerate ground states. The presence of these emerging low energy doublet degrees of freedom necessitates the consideration of inter-tetrahedron interactions, even in limit of large $d'/d$ breathing ratio. For example, how this degeneracy is resolved by the collective physics of the tetrahedra and the nature of the ultimate ground state of \byzo{} remain open questions.
 
Turning now our focus to \byzo{}, we begin with the single-ion physics. The magnetism of \byzo{} originates from the Yb$^{3+}$ ions where the $J=7/2$ manifold of the isolated ion is split by the $C_{3v}$ symmetric crystal field. This crystal field leads to a Kramers doublet separated by a gap of $38.2$ meV from the remaining levels~\cite{HakuJPSJ2016}. Given this splitting is large relative to the scale of the ion-ion interactions~\cite{RauPRB2018}, a description in terms of an effective $S = 1/2$ degree of freedom for each Yb$^{3+}$ ion is possible~\cite{RauGingrasAnnuRevCMP,RauPRB2018}. Analysis of inelastic neutron scattering data on powder samples found that antiferromagnetic Heisenberg exchange is dominant~\cite{KimuraPRB2014} with a surprisingly strong, but sub-dominant, Dzyaloshinskii-Moriya interaction~\cite{RauPRL2016,HakuPRB2016,Park2016} within each small tetrahedron.  The absence of magnetic ordering down to $\sim 0.1$ K~\cite{HakuPRB2016,RauJPCM_2018}, despite a Curie-Weiss temperature of $-6.1$ K (as measured in the present paper), the observation of a residual entropy of ~$k_{\mathrm{B}}\ln(2)/4$ per Yb~\cite{KimuraPRB2014} and a dispersionless excitation spectrum~\cite{RauPRL2016,HakuPRB2016,Park2016} is consistent with these smaller tetrahedra remaining mostly decoupled 
(i.e. paramagnetically) down to $T \sim $ 0.1 K.

The unexpected observation of a robust Heisenberg plus Dzyaloshinskii-Moriya interactions, with negligible symmetric anisotropies between the effective $S=1/2$ Yb$^{3+}$ spins in \byzo{}~\cite{RauPRB2018}, 
has led to new insights in understanding related ytterbium compounds such as the chalcogenide spinels~\cite{spinel1,spinel2,spinel3} AYb$_2$X$_4$ (with A = Cd, Mg and X = S, Se) and the nearly perfect $S=1/2$ honeycomb  Heisenberg antiferromagnet YbCl$_{3}$~\cite{ybcl31,ybcl32,ybcl33}.

Proximity to this regime of weak symmetric exchange anisotropy in breathing pyrochlores with a large breathing ratio ($d'/d \gg 1$) has also been predicted to lead to rich collective physics when these tetrahedra are weakly coupled. Theoretical proposals include realizations of bosonic analogues of topological band structures such as Weyl touching points~\cite{Li2016}, enhanced stability of quantum spin ice~\cite{SavaryPRB2016,Gingras_2014}, and possibly the emergence of a rank-2 $U(1)$ spin liquid~\cite{YanPRL2020}. Realization of these kinds of novel collective physics of the small tetrahedra requires inter-tetrahedron interactions to be at play, and thus their presence and magnitude to be
exposed and characterized experimentally. 

For the bulk of the experimental data on \byzo{} thus far collected, these have been found to be too small to be observable, with a model of isolated tetrahedra proving sufficient to reproduce qualitatively all of the experimental data above $\sim 1$ K~\cite{RauPRL2016,HakuPRB2016,Park2016,RauJPCM_2018}. However, notable discrepancies appear in the heat capacity~\cite{RauPRL2016,HakuPRB2016,RauJPCM_2018} below $\sim 0.3$ K, suggesting the single-tetrahedron model is incomplete at very low energies. Inter-tetrahedron interactions could provide a natural explanation~\cite{RauPRL2016}, though the effects of structural disorder~\cite{HakuPRB2016} must also be considered carefully, given the non-Kramers $E$ doublet ground state of each small tetrahedron~\cite{RauPRL2016,RauJPCM_2018} can be split by non-magnetic disorder. Resolving the origin of these discrepancies through the investigation of
powder samples is difficult, not only due to the small energy scales involved, but also due to the directional averaging inherent for
powders~\cite{RauJPCM_2018}. In particular, in neutron scattering, this averaging can render dispersive excitations difficult to resolve beyond a linewidth. 
Thus, to explore carefully the low-temperature physics of \byzo{} and understand the nature of any inter-tetrahedron interactions, single-crystal studies are indeed required. Moreover, single-crystal-studies allow access to the details of field-driven level crossings even at the single-tetrahedron level which are unresolvable using powder samples.

In this work, we report the synthesis and characterization results obtained for the first single-crystal samples of \byzo{}. We present comprehensive studies of the single-crystal \byzo{} using a wide range of experimental techniques including inelastic neutron scattering (INS) and bulk magnetic susceptibility, as well as ultra-sensitive tunnel diode oscillator (TDO) magnetic susceptibility measurements.  Our single crystal measurements allow us to directly probe the finite wave-vector dependence of the magnetic excitations and anisotropic response at low temperatures and high magnetic fields. We find that the single tetrahedron model maintains its qualitative agreement with inelastic neutron measurements for most magnetic fields. On the other hand, our diffuse neutron scattering data collected at 70 mK shows evidence of deviations from the single tetrahedron model near the critical magnetic field attributed to the first level crossing and falls short of explaining the very low energy behavior. These results suggest that the inter-tetrahedron interactions could be responsible for the low-energy and low temperature physics, as depicted, for example, by the diffuse neutron scattering above the critical field of the first level crossing in this system. These results are also in agreement with the previously reported low temperature specific heat measurements where the simple single tetrahedron model fails to reproduce the observed data at low temperatures~\cite{RauJPCM_2018}. This work paves the way for further theoretical and experimental investigations aimed at shedding light on the unsolved ground state of this interesting breathing pyrochlore system.

\section{EXPERIMENTAL AND THEORETICAL METHODS}
\label{sec:Methods}

%%%%%%%%%%%%%%%%%%%%%%%%%%%%%%FIG1%%%%%%%%%%%%%%%%%%
\begin{figure*}[htp]
	\centering
	\includegraphics[width=0.95\textwidth]{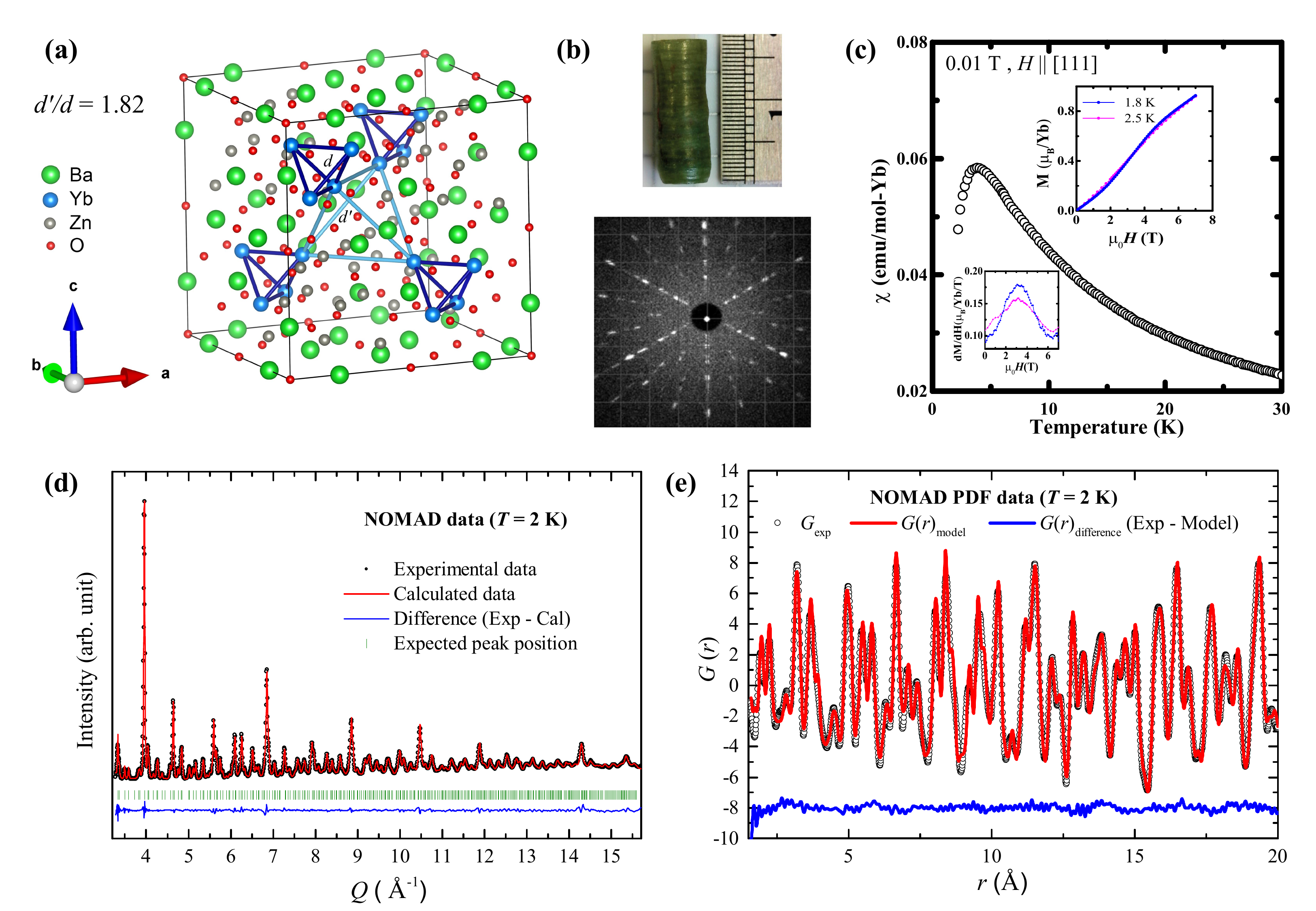}
	\caption{(a) The crystal structure of \byzo{}.  (b) Single crystal sample of \byzo{} used for single crystal neutron scattering experiment and the Laue back-scattering diffraction pattern of plane perpendicular to the $[111]$ direction. (c) Magnetic susceptibility $\chi$ vs temperature, $T$, down to 1.8 K. (Upper Inset) Field-dependent magnetization $M(H)$ at different $T$, with $H$ sweep rate of 0.03 T/min. The inflexion point in the $M$ vs $H$ data  corresponds to a transition from an approximate total spin $S_{{\rm tot}}$ = 0 state to an approximate $S_{{\rm tot}}$ = 1 state. (Lower Inset) First derivative of magnetization $M(H)$ as function of field $H$ at temperatures of $T=1.8$ K and $T=2.5$ K. (d) The Rietveld refinement of the neutron powder diffraction pattern of \byzo{} at 2 K collected using NOMAD at the Spallation Neutron Source of
ORNL. The red and blue solid lines represent the calculated
intensity and the difference between
observed and calculated intensities (weighted profile $R$-factor~\cite{R_Factors}, $R_{\rm wp}$ = 4.79\%), while the short vertical green marks represent the expected Bragg peak positions. (e) The pair density function (PDF) data at 2 K along with the refinement is shown. Black open circles, red lines and blue lines represent the PDF data,  the refinement using the PDFGUI software~\cite{Pdfgui} with assumed space group $F\bar{4} 3 m$ and the difference between PDF data and refinement, respectively.
		\label{fig:struct-bulk}
	}
\end{figure*}
%%%%%%%%%%%%%%%%%%%%%%%%%%%%%%%%%%%%%%%%%%%%%%%%%%%

\subsection{Sample Synthesis}
A powder sample of \byzo{} [structure shown in Fig.~\subref{fig:struct-bulk}{(a)}] was synthesized, as reported earlier \cite{KimuraPRB2014}, by finely grinding mixed powders of Yb$_2$O$_3$ (99.99 \%, Alpha Aesar), ZnO (99.99 \% , Sigma Aldrich) , and BaCO$_3$ (99.95 \%, Alpha Aesar) in the stoichiometric ratio of 1:5:3 and reacting at 1140~$^{\circ}$C in a box furnace for 48 hours with several intermediate grindings at a lower temperature.  
In preparation for the single crystal growth, the powder was compressed hydrostatically into a cylindrical rod, and then sintered at 1150~$^{\circ}$C in a vertical Bridgman furnace.  
Finally, large single crystals of \byzo{} were grown using the optical floating zone method. A typical crystal was grown in an O\textsubscript{2} atmosphere at 0.9 MPa, with an initial growth speed of 10 mm/hr and, upon stabilization of the liquid zone, at $\sim 6$ mm/hr until finished. To ensure uniform homogeneity of the liquid zone, the feed and seed rods were rotated in opposite directions with a rotation speed of 20 rpm during growth.

\subsection{Susceptibility and Magnetization measurements}
\label{sec:Susceptibility and Magnetization}

Magnetic susceptibility, $\chi$, and magnetization, $M$,  measurements were performed on a single crystal sample of \byzo{} up to $\mu_0H=$~7 T, using an in-house Cryogenic S700X SQUID magnetometer at temperatures of 1.8 K to 200 K using a Helium-4 probe. To explore the high-field magnetic properties of the \byzo{} crystals, tunnel diode oscillator (TDO) measurements (described below) \cite{TDOReference} were carried out at the DC Field Facility of the National High Magnetic Field Laboratory in Tallahassee. A dilution refrigerator and a $ ^{3} $He system were used to cover the temperature range from 41 mK to 20 K. The field-dependent measurements up to 18 T were conducted using a superconducting magnet. The field sweep rate was kept low (0.1 to 0.3 T/min)  to minimize the magnetocaloric effect. For a typical TDO measurement, bar-shaped \byzo{} crystals of $\sim$ 2 mm in length and $\sim$ 1 mm in transverse width were prepared and placed inside a detection coil, with the $[111]$ direction aligned along the coil axis. Together, the coil and sample within form the inductive component of a LC circuit. The LC circuit, powered by a tunnel diode operating in its negative resistance region, was tuned to resonance in a frequency range between 10 and 50 MHz. The shift in the resonance frequency $f$, which is related to the change in the sample magnetization $M$ ($df/dH \propto d^2M/dH^2$)~\cite{Shi2019}, was then recorded. The TDO method measures the resonance frequency to a very high precision \cite{TDOReference} which enables identifying changes in the magnetic moments down to $\sim$ $10^{-12}$ emu compared to $\sim$ $10^{-8}$ emu in SQUID magnetometry measurements.

\subsection{Neutron Scattering}
\label{sec:Neutron Scattering}
Inelastic neutron scattering (INS) measurements were performed at a number of facilities:  using the Disk Chopper Spectrometer (DCS) at the National Institute of Standards and Technology, the Cold Neutron Chopper Spectrometer (CNCS) at Oak Ridge National Laboratory and the Wide Angle Neutron Diffractometer (WAND$^2$) also at Oak Ridge National Laboratory. 
A superconducting magnet was used in each of the above instrument to provide, in all cases, a vertical magnetic field up to 10 T (DCS), 8 T (CNCS) and 5 T (WAND$^2$). In all experiments, a single crystal sample of mass of about 1 g was mounted on a Cu sample holder in a dilution refrigerator. At DCS and WAND$^2$, the sample was mounted with the $[{h+k},-{h}+k,-2k]$  scattering plane being horizontal and with the magnetic field applied along the $[111]$ cubic direction. Neutron-absorbing Cd was used to shield the sample holder to reduce background scattering. The experiment at CNCS was performed with the sample mounted so the horizontal scattering plane is $[{h}{h}l]$ with the magnetic field applied vertically, along the $[1\bar{1}0]$ cubic direction. In all experiments,  low temperature measurements were conducted with applied fields at the
 70 mK base temperature and the background determination measurements were conducted under zero magnetic field at 50 K, 100 K and 50 K for DCS, CNCS and WAND$^2$, respectively. The sample stage was rotated close to 180$^\circ$ to cover few  Brillouin zones in all the experiments.
 After carefully investigating the ``as-collected'' (unsymmetrized data), we used the crystallographically-allowed  symmetry operations to improve statistics. 
The specific symmetry operations applied for each figure reported below 
in Sec.~\ref{sec:neutron} are noted in the pertinent figure captions.  
In order to investigate any structural disorder in \byzo{}, total neutron scattering measurements were performed on a polycrystalline sample using the Nanoscale-Ordered Materials Diffractometer (NOMAD) at the Spallation Neutron Source (SNS) at Oak Ridge National Laboratory (ORNL, USA) with a lowest temperature of 2 K considered. 
Analysis and visualization of the neutron scattering data were performed using DAVE MSlice~\cite{DAVEMSLICE}, Mantid ~\cite{Mantid} and Python software packages.

\subsection{Single Tetrahedron Model}
\label{sec:singleTetra}

Given the large inter-tetrahedron distance of $d' \gtrsim 6$ \AA{} compared to the small intra-tetrahedron distance $d \sim 3.3$ \AA{} [as shown in Fig.~\subref{fig:struct-bulk}{(a)}], to model in \byzo{}, we begin by assuming that all inter-tetrahedron interactions are negligible and model the system as a set of decoupled “small” tetrahedra, hence focusing on the intra-tetrahedron interactions, which are expected to dominate. Since the first excited crystal field level is $\sim 38.2 \meV$ above the ground doublet~\cite{HakuPRB2016}, we can project
the Hamiltonian describing the Yb$^{3+}$--Yb$^{3+}$ interactions~\cite{RauPRB2018}
into the lowest ground state doublet, ignoring any higher-order perturbative corrections from virtual crystal-field excitations~\cite{RauGingrasAnnuRevCMP,RauPRB2018,PhysRevB.93.184408,Molavian}. Symmetry strongly constrains the allowed exchange interactions between these resulting (from the projection) 
$S=1/2$ effective spins, with a bond symmetry group of $2mm$ and full tetrahedral symmetry $\bar{4}3m$ about the center of each tetrahedron~\cite{inta}.
These interactions are expected to act pair-wise between the effective Yb$^{3+}$ spins $1/2$  in each tetrahedron~\cite{RauPRB2018}, $\vec{S}_i$, with the model for each individual tetrahedron taking the form~\cite{RauPRL2016,Park2016,HakuPRB2016}
\begin{align}
\label{eq:model}
{\cal H}_{\rm eff} \equiv &\sum_{i=1}^4 \sum_{j<i} \left[
J_{zz} S^z_i S^z_j - J_{\pm}\left(S^+_i S^-_j+S^-_i S^+_j\right)+\right.  \nonumber \\
& J_{\pm\pm} \left(\gamma_{ij} S^+_i S^+_j+\hc \right)+ \\
& \left. J_{z\pm}  \left(
\zeta_{ij} \left[ S^z_i S^+_j+ S^+_i S^z_j \right]+ \hc \right) \nonumber\right] -\muB  \sum_{i=1}^4   \vec{B} \cdot \vec{\mu}_i ,
\end{align}
where we have included a magnetic field $\vec{B}$ with the local Yb$^{3+}$ magnetic moment operators $\vec{\mu}_i$ given by
\begin{equation}
\label{eq:moment}
\vec{\mu}_i \equiv  \mu_B \left[ g_{\pm} \left(\vhat{x}_i S^x_i +  \vhat{y}_i S^y_i\right) + g_z \vhat{z}_i S^z_i\right] ,
\end{equation}
 and associated  $g$-factors, $g_z$ and $g_{\pm}$.
We borrow definitions from Ref.~\cite{RossPRX2011} for the bond form factors $\gamma_{ij}$ and $\zeta_{ij}$ and the local axes ($\vhat{x}_i$, $\vhat{y}_i$, $\vhat{z}_i$). Note that there is freedom in choosing the sign of $J_{z\pm}$, as it can be changed by a basis transformation.

The spectrum of this Hamiltonian is partly constrained by tetrahedral symmetry. 
The sixteen states of the four-spin system break into the irreducible representations of $\bar{4}3m$
\begin{equation}
\label{eq:IR}
A_2 \oplus 3 E\oplus T_1 \oplus 2 T_2,
\end{equation}
under the action of the tetrahedral group~\cite{RauJPCM_2018}. 
This gives a level structure that consists of a singlet ($A_2$), three doublets ($E$) and three triplets ($T_1$ or $T_2$),with the highest lying doublet ($E$) and triplet ($T_2$) being nearly degenerate; see 
Fig.~\subref{fig:TDO_hightemp}{(a)}.
We use the parameters of Rau \textit{et al.}~\cite{RauJPCM_2018}, which were fit using data from powder samples in finite magnetic fields, and do not reconsider here these values:
\begin{align}
\label{eq:JValues}
J_{zz} &= -0.040 \meV, &	J_{\pm} &= +0.141 \meV ,\nonumber\\	
J_{\pm\pm} &= +0.160 \meV, & 	J_{z\pm}  &= +0.302 \meV ,
\end{align}
with $g$-factors  $g_z$ = 2.726 and $g_{\pm}$ = 2.301.

Theoretically, we compute the unpolarized inelastic neutron scattering intensity, $I(\vec{Q},E)$, using the same methods presented in Refs.~[\onlinecite{RauPRL2016},\onlinecite{RauJPCM_2018}], which is given by
\begin{align}
\label{eq:NeutronIntensity1}
I(\vec{Q},E)&\propto F(\vec{Q})^2\sum_{\mu \nu}(\delta_{\mu \nu}-\hat{Q}_\mu \hat{Q}_\nu)S_{\mu \nu}(\vec{Q},E) ,
\end{align}
where $F(\vec{Q})$ is the atomic form factor of Yb\textsuperscript{3+}~\cite{intc}. The spin structure factor for each tetrahedron is given by
\begin{align*}
%\label{eq:NeutronSpinStructureFact}
S_{\mu \nu}(\vec{Q},E)\equiv \sum_{nm} \rho_{m} \braket{m | \mu^\mu_{-\vec{Q}} | n}\braket{n | \mu^\nu_{\vec{Q}} | m}\delta(E-E_n+E_m),
\end{align*}
where $\rho_m\equiv e^{-\beta E_m/T}/Z$ is the Boltzmann weight and the Fourier transform of the magnetic moment operator is given by $\vec{\mu}_{\vec{Q}}\equiv {\frac{1}{4}}\sum_{i=1}^4 e^{i\vec{Q} \cdot \vec{r}} \vec{\mu}_i$, with $\vec{\mu}_i$ given by Eq. (2).

\section{RESULTS AND DISCUSSION}
\label{sec:Results}

\subsection{Magnetic property measurements}
\label{sec:TDO}

The magnetic susceptibility, $\chi$, was measured on our single crystal sample down to 1.8 K under an applied magnetic field of 0.01 T along the crystallographic $[111]$ direction using the aforementioned Cryogenic S700X SQUID system, as shown in Fig.~\subref{fig:struct-bulk}{(c)}. A broad maximum is observed at $\sim$4 K, similar to that in the powder sample~\cite{KimuraPRB2014}. The Curie-Weiss temperature $\theta_{{\rm CW}} = -6.85$ K obtained from the  \byzo{} single crystal susceptibility data by fitting $\chi = C/ (T -\theta_{{\rm CW}})$ in the temperature range of 10 K $<T<$30 K,  agrees well with the previously reported value $\theta_{{\rm CW}} = -6.7$ K using a powder sample \cite{KimuraPRB2014}. 

%%%%%%%%%%%%%%%%%%%%%%%%%%%%%%FIG2%%%%%%%%%%%%%%%%%%
\begin{figure*}[htp]
	\centering
	\includegraphics[width=0.95\textwidth]{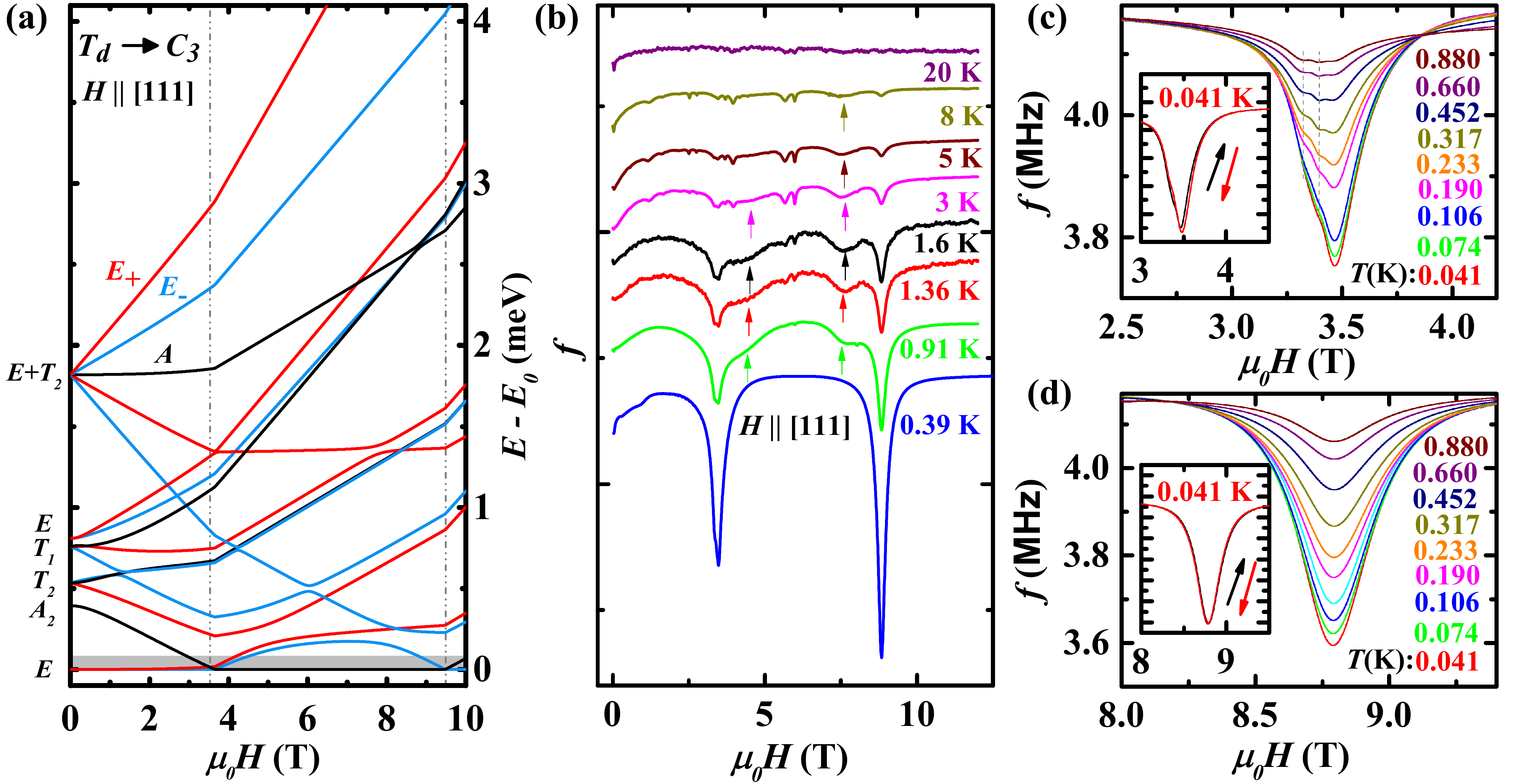}
	\caption{(a) Field-dependent energy levels calculated by the single-tetrahedron model. Irreducible representations of the zero-field ($T_d$) and finite field ($C_3$) symmetry groups are indicated. Critical fields where  there is a ground state level crossing ($\mu_{0}H_{\rm{c}1} $ = 3.65 T and $\mu_{0}H_{\rm{c}2} $ = 9.48 T) indicated by vertical dash lines are close to the experimental values. The low energy regime (e.g. as would be thermally populated at $ T\lesssim 1$ K) is indicated by a gray shaded box.  (b)  Tunnel diode oscillator (TDO) frequency ($f$) as a function of magnetic field $\mu_{0}H$ $(|| ~[111])$ up to 20 K. Traces are shifted vertically for clarity. Small dips in frequency in addition to two level crossings are marked by arrows.
	The sharp features at higher temperatures, e.g. near $\sim 6$ T, are associated with background signal from our experimental apparatus.
	(c,d) $f$ vs $\mu_{0}H$ $(||~ [111])$ as a function of magnetic field $\mu_{0}H$ $(|| ~[111])$ up to 0.880 K, near the ground state level crossings at (c) 3.5 T, and at (d) 8.8 T. Additional features observed near 3.5 T are marked by grey dash lines. Insets in (c) and (d) show $f$ vs $\mu_{0}H$ for the upsweep (black) and downsweep (red) at $T$ = 0.041 K, where no hysteresis is seen. 	\label{fig:TDO_hightemp}
     }
\end{figure*}
%%%%%%%%%%%%%%%%%%%%%%%%%%%%%%%%%%%%%%%%%%%%%%%%%%%

Applying larger magnetic fields, \byzo{} can be tuned through two transitions. In powder samples, these appear near the critical fields $\mu_{0}H_{\rm{c}1}\approx 3.5\T$ and $\mu_{0}H_{\rm{c}2} \approx 8.8\T$~\cite{HakuPRB2016,Park2016}. We show in Fig. \subref{fig:TDO_hightemp}{(a)} the energy level results calculated using the single tetrahedron model as a function of field. The theoretical values of $\mu_{0}H_{\rm{c}1}$ and $\mu_{0}H_{\rm{c}2}$ are indicated by vertical dash lines.
As alluded to earlier, the TDO frequency $f$ is an extremely sensitive measurement of the magnetic susceptibility~\cite{Haravifard2016,Shi2019,Steinhardt2021} and reveals fine details of changes in the magnetic response of sample. 
We show in Fig.~\subref{fig:TDO_hightemp}{(b-d)} the measured TDO frequency in \byzo{} over a range of magnetic fields along the $[111]$ direction, with strength $0\T \leq \mu_0 H \lesssim 12\T$. 
Our SQUID and TDO results collected on the single crystal reveal that the critical fields for $H \parallel [111]$ [Fig. \subref{fig:TDO_hightemp}{(c)} and \subref{fig:TDO_hightemp}{(d)}] are nearly identical to those reported previously on powder samples~\cite{Park2016}.

As shown in Fig.~\subref{fig:TDO_hightemp}{(c)}, the TDO data reveals the presence of two shoulder-like structures for field values just below the first level crossing at 3.5 T and at the lowest temperature, and become more distinguishable as temperature is increased. These features may suggest the presence of additional level crossings near $\mu_{0}H_{\rm{c}1}$ that are not captured by the single-tetrahedron model. We leave a detailed exploration of the nature of these features to future work.  Similar to our SQUID data, we do not see obvious hysteresis either near $\mu_{0}H_{\rm{c}1} $ or $\mu_{0}H_{\rm{c}2} $. This is in contrast to Ref.~\cite{Park2016}, where a field-sweep-rate dependent hysteresis was reported on powder samples of \byzo{} and attributed to an avoided level crossing of the ground and first excited state near $\mu_{0}H_{\rm{c}1}$.~\footnote{Depending on the magnitude of the field sweep rate relative to the energy splitting at the avoid crossing, non-adiabatic transitions between the ground and first excited states are possible and their probability can be estimated using the usual Landau-Zener formula. Significant non-adiabatic probability implies hysteresis in the state of the system (and thus magnetization) as the field is swept up or down.}
 However, in a $[111]$ field, the crossing at $\mu_0 H_{\rm{c}1}$ is \emph{not avoided}; symmetry forbids any coupling between these states [see Fig.~\subref{fig:TDO_hightemp}{(a)}]. 
 We note that no hysteresis is observed (or is in fact expected~\footnote{
 The authors of Ref.~\cite{Park2016} attribute the hysteresis observed in their data to a non-equilibrium effect due to the finite rate of change in the magnetic field strength. However, estimates of the probability of non-adiabatic transitions (using the standard Laudau-Zener formula~\cite{landau1987quantum}) for any reasonable splitting size [$O({\rm meV})$] and field sweep rate [$O({\rm mT/min}]$] are negligible.
 }) in our data, in contrast to that seen in Ref.~\cite{Park2016} [insets of Fig.~\subref{fig:TDO_hightemp}{(c)} and ~\subref{fig:TDO_hightemp}{(d)}]. With increasing temperature to 20 K [Fig. \subref{fig:TDO_hightemp}{(b)}], the sharp features associated with the two ground state level crossings at $\mu_{0}H_{\rm{c}1} $ and $\mu_{0}H_{\rm{c}2} $ get broader and diminished, while new features as indicated by arrows, appear as the temperature is raised.

From measurements on our single crystal, we have exposed signatures of two field-driven transitions as expected from the theoretical  single-tetrahedron model. We next consider how the spin-spin correlations, as probed by neutron scattering measurements on our single crystal, evolve as a function of field and, in particular, across the two transitions at $\mu_{0}H_{\rm{c}1} $ and $\mu_{0}H_{\rm{c}2}$.

\subsection{Neutron scattering}
\label{sec:neutron}
In this section, we present the results of three complementary neutron scattering experiments. In order, these address the presence or absence of structural disorder (Sec.~\ref{subsec:neutron-struct}), the finite-energy magnetic excitations (Sec.~\ref{subsec:neutron-inelastic}) as well as an indirect view on the low-energy excitations via an energy-integrated experiment (Sec.~\ref{subsec:neutron-diffuse}).

\subsubsection{Structural characterization}
\label{subsec:neutron-struct}

One of the main concerns in previous studies on \byzo{}~\cite{HakuPRB2016,RauPRL2016,RauJPCM_2018} was whether or not any disorder played a role in the discrepancies observed in their reported low temperature measurements relative to the theoretical results on the above single tetrahedron model, especially in the heat capacity data. In an attempt to address this concern and investigate the detailed crystal structure at low temperature and any possible disorder, neutron scattering measurements were performed on \byzo{} powder sample using the Nanoscale-Ordered Materials Diffractometer (NOMAD) at Oak Ridge National Laboratory. The data was collected for selected temperatures between 100 K to 2 K. Rietveld refinements were performed using the GSAS II software on the collected data. No structural transition is observed down to 2 K and the system can be well described with the $F\bar{4} 3 m$ space group, as shown in Fig 1(d). To understand the local atomic structure of \byzo{}, the pair distribution function (PDF) of the total scattering data was also analyzed \cite{TAKESHI201255}. The PDF data, $G(r)$, collected at 2 K is shown in Fig. 1(e). Data were refined using the PDFGUI software ~\cite{Pdfgui} with the $F\bar{4} 3 m$  space group. As evident from the refinement, the observed data is well represented by our model with $F\bar{4} 3 m$ symmetry with weighted $R$-factor~\cite{R_Factors} $R_{\rm w}=5\%$. No additional peak splitting or significant mismatch is resolvable; our measurements thus do not provide any evidence for local structural disorder in this system down to 2 K.

%%%%%%%%%%%%%%%%%%%%%%%%%%%%%%FIG3%%%%%%%%%%%%%%%%%%

\begin{figure}[htp]
	\centering
	\includegraphics[width=0.46\textwidth]{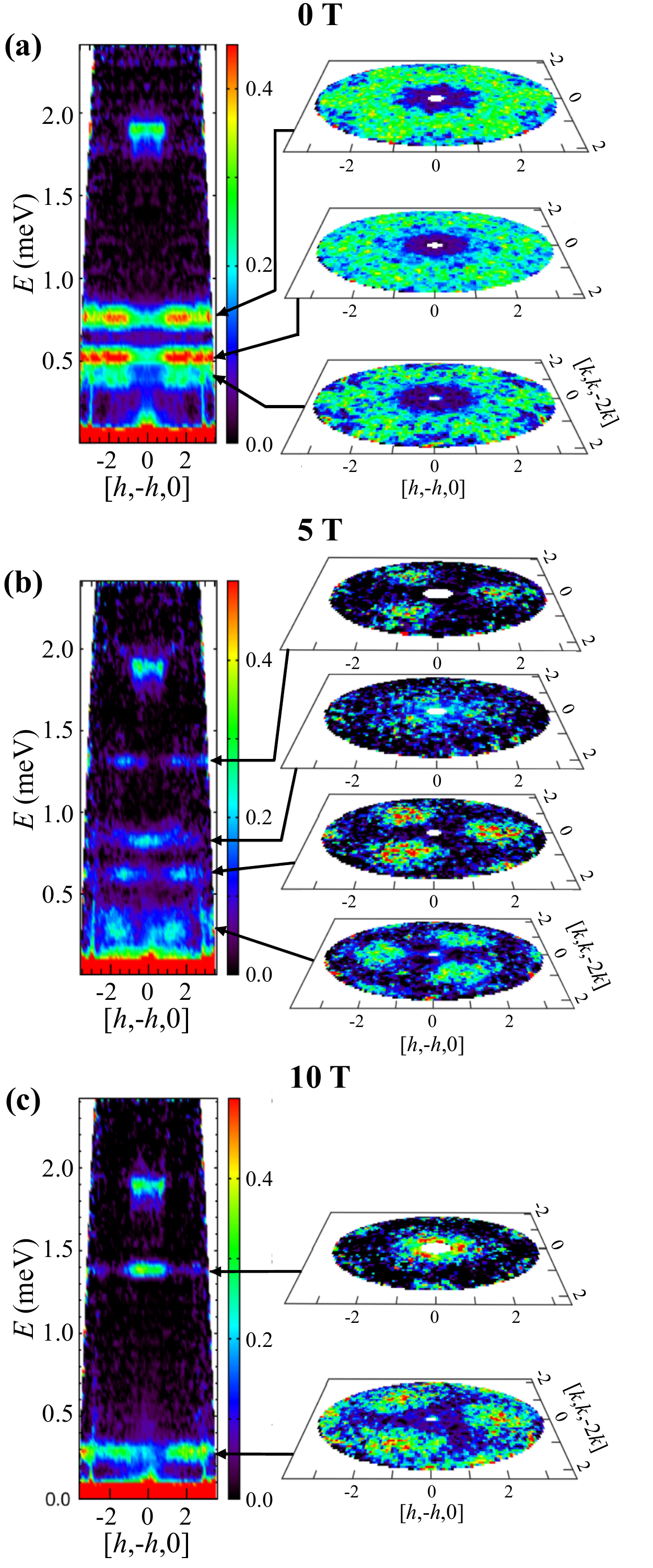}
	\caption{Intensity for inelastic neutron scattering from a single crystal sample of \byzo{} measured at $T$ = 70 mK using the  Disk Chopper Spectrometer (DCS) for a series of  fields  $\vec{H}\parallel[111]$ (a) 0 T (b) 5 T and (c) 10 T. For each field, the left panel shows the intensity as a function of energy transfer and momentum along $[h,-h,0]$ direction. On the right, constant energy intensity maps in the $[h+k,-h+k,-2k]$ plane are shown for selected dispersionless bands which are integrated along energy with an energy width specified to fully capture each band for each magnetic field.
		\label{fig:Neutron_SQW}
	}
\end{figure}
%%%%%%%%%%%%%%%%%%%%%%%%%%%%%%%%%%%%%%%%%%%%%%%%%%
\subsubsection{Energy-resolved inelastic scattering}
\label{subsec:neutron-inelastic}

To date, only powder samples of \byzo{} have been measured with inelastic neutron scattering~\cite{HakuPRB2016,RauPRL2016,RauJPCM_2018} and, within the constraints of powder averaged inelastic neutron scattering data, the experimentally observed low energy excitations are in agreement with the single tetrahedron model~\cite{RauPRL2016,HakuPRB2016,Park2016}. In order to study the detailed physics of this system beyond the single tetrahedron model, such as the details of the energy levels crossing depending on the direction of applied field, single crystal neutron scattering experiments performed at zero and finite fields (applied along different lattice directions) are necessary. 
Single crystal samples also allow us to investigate the anisotropic behavior of the excitations compared to previous powder neutron scattering studies. 

To this end, we have performed inelastic neutron scattering experiments using our single crystal sample of \byzo{} as explained in Sec.~\ref{sec:Neutron Scattering} with applied field $\vec{H} \parallel [111]$ (at DCS) and $\vec{H} \parallel [1\bar{1}0]$ (at CNCS).
The field evolution of the low energy excitations for a few selected fields between $\mu_{0}H_{\rm{c}1} $ and $\mu_{0}H_{\rm{c}2}$ are shown in Fig.~\ref{fig:Neutron_SQW} with applied field $\vec{H} \parallel [111]$. 
Figure~\ref{fig:Neutron_SQW} shows the wave vector dependence for each energy band as a function of field, in which the cuts were performed over a narrow energy window. 

Similar to powder samples~\cite{RauPRL2016,Park2016,HakuPRB2016}, dispersionless excitations are observed at zero and finite fields in our single-crystal samples as well. 
A slight broadening of the excitations, greater than the resolution of the instruments, is observed, although no clear dispersion visible. We note that some intrinsic broadening of the levels is to be expected due to any dispersion-induced by inter-tetrahedron interactions; the slightness of this broadening is consistent with the smallness of these couplings.

%%%%%%%%%%%%%%%%%%%%%%%%%%%%%%FIG4%%%%%%%%%%%%%%%%%%
\begin{figure*}[htp]
	\centering
	\includegraphics[width=0.95\textwidth]{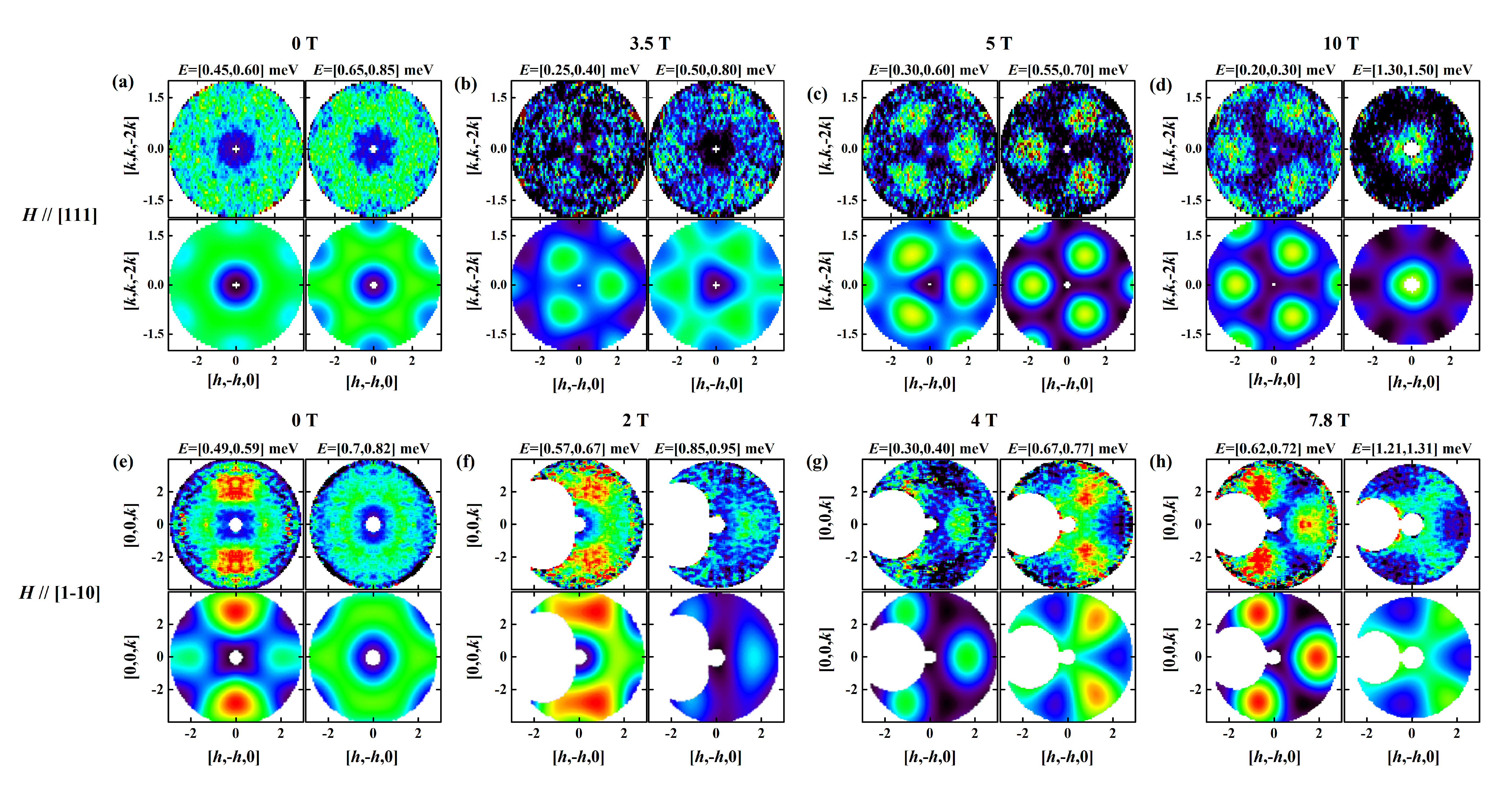}
	\caption{Comparison of the experimental (top row of each panel) and theoretical (bottom row of each panel) dynamical structure factor for the model of Eq. (1) with the parameters of Eq. (4) for several energy windows at different magnetic fields, including the atomic form factor of Yb$^{3+}$. The energy windows are convoluted with a finite Gaussian to emulate experimental broadening and resolution. In the following, experimental plots symmetry operations were applied to improve the statistics. $\vec{H}\parallel[111]$ at 0 T: 6-fold rotation and ~$H>$~0 T: 3-fold rotation. $\vec{H}\parallel[1\bar{1}0]$ at 0 T: mirror symmetry operations centered at $[h,-h,0]$ and $[0,0,k]$ directions and $H>0$ T mirror symmetry operations centered at $[0,0,k]$. $\vec{H}\parallel[111]$ data was measured at DCS with incident neutron energy $E_{\textrm i}$= 5 meV and only the detector background subtracted (i.e. not including the high-temperature background). $\vec{H}\parallel[1\bar{1}0]$ data was measured at CNCS and 100 K data was used as the background to eliminate unavoidable nearly temperature independent background scattering observed due to scattering from the magnet.
		\label{fig:Neutron_EnergyCuts}
	}
\end{figure*}
%%%%%%%%%%%%%%%%%%%%%%%%%%%%%%%%%%%%%%%%%%%%%%%%%%

Having access to single-crystal samples, we can explore the wave-vector dependence of the intensity of each dispersionless mode in 3-dimension reciprocal space and check for the theoretically expected symmetry of the neutron scattering intensity in the scattering plane in zero and finite fields.  In zero field, the presence of full $T_d$ symmetry and time-reversal symmetry imposes a six-fold symmetry
in the $[h + k, -h + k, -2k]$ plane. Thus, each band exhibits a pinwheel-type pattern which is 6-fold symmetric in this 
%$[h + k, -h + k, -2k]$ 
scattering plane, as shown in the three right panels of ig.~\subref{fig:Neutron_SQW}{(a)}.
Upon application of a $[111]$ field, we observe an evolution of the intensity as a function of wave-vector which changes qualitatively across each critical field (level crossing). The details of the redistribution of intensity depend on the field direction: for $[111]$ fields, only a three-fold symmetry remains, while for fields along $[1\bar{1}0]$, only a single two-fold symmetry is present. The zero-field Hamiltonian ${\cal H}_{\rm eff}$ in Eq.~(\ref{eq:model}) has a mirror symmetry, $\sigma = - \vec{C}_2$, that is broken by finite field. However, this mirror reverses a $[111]$ magnetic field and therefore, combining it with time-reversal, yields an additional symmetry
\begin{align}
\label{eq:NeutronIntensityField}
I(\vec{Q},E;\vec{B})=I(\vec{C}_2 \vec{Q},E;\vec{B}),
\end{align}
for the $[111]$ field direction.

Thus, we effectively have a set of two-fold axes in the plane perpendicular to $[111]$, giving a symmetry group of $D_3$ rather than only a $C_3$ symmetry at zero field. Under a finite magnetic field, this effective inversion symmetry [Eq.~(\ref{eq:NeutronIntensityField})] is broken and the intensity pattern becomes 3-fold symmetric. This can be seen in Figs.~\subref{fig:Neutron_SQW}{(b,c)}, which show the $E$-$\vec{Q}$ plots and the energy-integrated cuts for each energy band for a $[111]$ field (5 T) above the first level crossing transition at $H_{\rm{c}1} $ and one (10 T) beyond the second level crossing transition at $H_{\rm{c}2} $. We show the results obtained for a 5 T applied field, where energy cuts obtained for the band centered at $E=0.25$ meV reveal broad features centered around $[20\bar{2}]$ that have the expected three-fold symmetry in the $[h+k,-h+k,-2k]$ plane. It is clear that the obtained pattern for 5 T is  different from the results collected in  zero field. For the next higher excitation band at $E= 0.65$ meV, the pattern is rotated by 30 degrees compared to the band at $E=0.25$ meV. Similarly, the intensity pattern in $\vec{Q}$ space in this scattering plane changes as we move up in the energy for each flat excitation band and shows a unique pattern for each. Similar to the 5 T results, the 10 T field data also demonstrates a subsequent change in the intensity following the second level crossing. For the energy transfer range determined by our incident energy, we could only observe two flat excitation bands at 10 T. The band centered at $E=0.25$ meV shows a similar 3-fold pattern for this field as observed for 5 T, however, the second band at $E=1.4$ meV shows an intensity that is more concentrated at small wave-vector; likely a transition between approximate $S_{\rm tot}=2$ states of mostly ferromagnetic character.

In order to explore the anisotropy of the neutron scattering intensity as function of the field direction, we also performed inelastic neutron scattering experiments with $\vec{H} \parallel [1\bar{1}0]$, perpendicular to the $[111]$ direction. As with the $[111]$ direction, we consider individual energy cuts at several fields ranging from 0 T to 10 T, comparing the applied field along $[1\bar{1}0]$ direction to the analogous cut for $[111]$ field. The results are shown in Fig.~\ref{fig:Neutron_EnergyCuts}. For each panel, the experimental data and theoretical calculation using single tetrahedron model are shown for both field directions, $\vec{H} \parallel [111]$ and $\vec{H} \parallel [1\bar{1}0]$.  
We emphasize that  Fig.~\subref{fig:Neutron_EnergyCuts} ~shows good agreement between the experimental data and the single tetrahedron model calculations, for all energy cuts and for both field directions, and within the constraints from experimental noise, background and resolution. All qualitative features for each energy cut are well captured by the theoretical calculations. This indicates that the nature of the dispersionless bands at energies $E>0.3$ meV are well explained by the single tetrahedron model. The consistency of the single-crystal data and the theoretical model fit to data from polycrystalline samples is non-trivial, given the different sample growth processes (powder vs. single crystal). Further, given the information lost due to powder averaging, one may worry that the theoretical model may not have been fully constrained by the polycrystalline data~\cite{RauPRL2016,RauJPCM_2018}. The  agreement reported here is evidence that the single-crystal and powder samples are exhibiting the same physics, and the model determined from powder samples remains valid, at least for energies $\gtrsim O(0.1 \meV)$.

\subsubsection{Energy-integrated diffuse scattering}
\label{subsec:neutron-diffuse}

\begin{figure}[tp]
	\centering
	\includegraphics[width=0.5\textwidth]{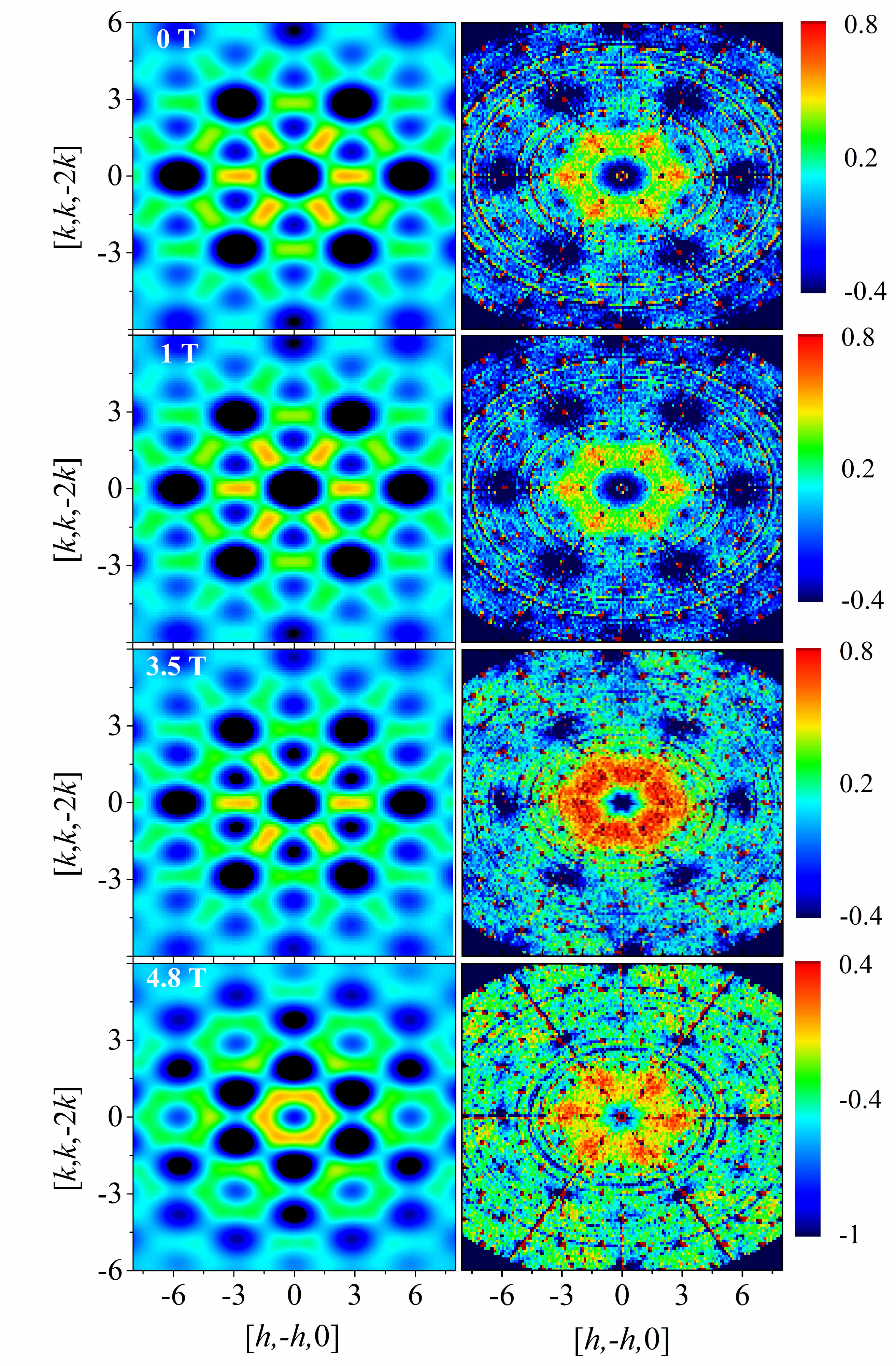}
	\caption{Comparison of the experimental (right) and  theoretical (left) energy-integrated structure factor at several magnetic fields using single tetrahedron model. Theoretical plots include the atomic form factor of Yb$^{3+}$. Experimental data was measured using the Wide Angle Neutron Diffractometer (WAND$^2$). Both experimental and theoretical plots show the data at $T$ = 70 mK after subtracting the background at 50 K at 0 T. 
	\label{fig:WANDdata}
	}
\end{figure}

While, the inelastic scattering data in Sec.~\ref{subsec:neutron-inelastic} provides detailed information about the finite-energy excitations, due to the challenges arising from experimental background, instrument resolution and the limited flux, it offers limited information about the physics of the system at low energies, where one might expect to look for evidence of physics beyond the single tetrahedron model.
Therefore, in order to capture this very low-energy physics, a diffuse neutron scattering experiment, which integrates over \emph{all} energies, was performed using the Wide Angle Neutron Diffractometer (WAND$^2$) instrument at Oak Ridge National Laboratory. WAND$^2$ features a highly efficient high resolution ${}^3$He 2-dimensional position sensitive detector which enables it to map a large portion of reciprocal space for single crystals. One of the specialized data collection purposes of WAND$^2$ is measurements of diffuse scattering in single crystals.
Figure ~\subref{fig:WANDdata}~ shows the neutron scattering data measured at WAND$^2$. Data was collected at 70 mK and fields up to 4.8 T. High temperature data collected at 50 K and zero field, was used as  background and subtracted from the 70 mK data for each field. The incident neutron wavelength at WAND$^2$ is $\lambda= 1.486$ \AA{}, with corresponding incident energy $E_{\textrm i}=37.04$ meV. Due to the inability to discriminate the energy and considering the energy resolution ($>3$ meV), one can expect the data obtained at  WAND$^2$ to amount to energy-integrated scattering up to $\sim 2$ meV. The diffuse scattering intensity (static structure factor) is given by integrating Eq.~(\ref{eq:NeutronIntensity1}) over energy
\begin{align}
\label{eq:IQE_integrated}
I(\vec{Q},B) \propto  F(\vec{Q})^2 \sum_{\mu \nu}(\delta_{\mu \nu}-\hat{Q}_\mu \hat{Q}_\nu)\braket{\mu^\mu_{-Q}\mu^\nu_{Q}}
\end{align}
where, again, $F(\vec{Q})$ is the atomic form factor of Yb\textsuperscript{3+}~\cite{intc}. Per its definition,  this is symmetric under $\vec{Q} \rightarrow -\vec{Q}$, even for finite magnetic fields for unpolarized neutron scattering intensity. Combining this with the three-fold rotational symmetry, the integrated intensity has an enlarged six-fold symmetry for a $[111]$ field.~\footnote{We note that this additional symmetry arises due to the symmetrization of the unpolarized intensity and the integration of over energy transfer that reduces the structure factor to static moment-moment correlation function.}

In zero field, the qualitative features of the data are reproduced in the calculation: a hexagon of high-intensity with maximum intensity along $[h,-h,0]$ and equivalent directions. In addition, it shows minima in the intensity at zero, as well as at six points equivalent to $\sim [6,-6,0]$. Six weaker minima are also visible near $[2,2,-4]$ (and equivalent positions) in both the theoretical calculation and the experimental data. However, the agreement is not perfect; these six minima are more pronounced in the theoretical calculation as compared to the experimental data. The pattern for $\mu_{0}H = 1$ T is similar to the zero field one in both the theoretical calculation and the experimental data. In contrast, calculated intensity at $\mu_{0}H = 3.5$ T is considerably different from the experimental data. Namely, the high-intensity hexagon is broader, with a more uniform intensity and minima near $[2,2,-4]$ and its equivalents are essentially absent. However, theoretical calculations obtained for 3.5 T, show very similar pattern to the 0 T and 1 T data, with same set of pronounced minima near $[2,2,-4]$.  At 4.8 T, the theoretical result differs from the lower field results. While some experimentally observed features, such as the location of intensity minima, agree with the theoretical results for the 4.8 T data, there are qualitative differences such as absence of intensity modulations in the high-intensity hexagon in the theoretical calculations, as well as in the visibility of the low intensity regions at $[6,-6,0]$ and equivalent positions. In comparison, the DCS and CNCS inelastic neutron scattering results, not at low energy,  as discussed in Sec.~\ref{subsec:neutron-inelastic},
were overall in acceptable agreement with the theoretical calculations using the single tetrahedron model (Figs.~\ref{fig:Neutron_EnergyCuts}). 

Recalling the aforementioned ability of the WAND$^2$ experiment to probe low-energy excitations, we interpret these results in the following way: as the inter-tetrahedron interactions are expected to be small, we can argue that the possible dispersion of the ground state doublet induced by these interactions is also small, with the corresponding excitations lying at very low energies, close to the elastic channel. The differences between the data and the theoretical calculations using the single tetrahedron model may thus be due to low energy diffuse scattering features arising from physics beyond the single tetrahedron model. Such discrepancies at low energy are consistent with the previously reported heat capacity data \cite{HakuPRB2016,RauJPCM_2018}, which showed a broad peak centered at $\sim$ 0.1 K, with an entropy change consistent with the release of two degrees of freedom per single tetrahedron. In the single-tetrahedron model, the ground state doublet is degenerate; the complete release of this entropy, indicating a unique ground state, thus points again to physics beyond the single-tetrahedron model. The energy scale of this splitting was estimated to be $\sim$ 0.015 meV at 0 T~\cite{HakuPRB2016} -- much smaller than the instrumental resolution of current neutron scattering experiments, $\sim$ 0.05 meV, as in the experiments presented in Sec.~\ref{subsec:neutron-inelastic}. We indicate this energy scale by a shaded gray box in Fig.~\subref{fig:TDO_hightemp}{(a)}. Compared to the single-tetrahedron energy levels, we see that this energy scale is indeed small.

%%%%%%%%%%%%%%%%%%%%%%%%%%%%%%FIG5%%%%%%%%%%%%%%%%%%
%%%%%%%%%%%%%%%%%%%%%%%%%%%%%%%%%%%%%%%%%%%%%%%%%%

The most natural origin for the splitting of the doublets may be attributed to inter-tetrahedron interactions or to non-magnetic disorder that lowers the symmetry of each tetrahedral cluster. Recalling Sec.~\ref{subsec:neutron-struct}, our PDF results show no observable evidence for disorder in our samples, indicating any non-magnetic structural disorder is  small and below the resolution threshold of our experiments. However, given the minute energy scale of the doublet splitting ($\sim 0.015$ meV), we cannot rule out that small distortions, below the experimental resolution, could potentially account for the entropy release.

At higher magnetic fields, away from the level crossings, the ground state is non-degenerate and thus the effect of either small inter-tetrahedron interactions or of non-magnetic disorder should be less significant. However, we note that there still exist major discrepancies, in the diffuse scattering data from WAND$^2$ as well as in the TDO measurements, with the single tetrahedron model at these higher fields. Crucially, both of these of measurements include contributions from the low-energy physics. A similar disagreement was noted in the high-field heat capacity~\cite{RauJPCM_2018}. The effective strength of the inter-tetrahedron interactions can vary between the different degenerate energy level manifolds of the tetrahedron and as a function of field as they mix. However, it is not clear that higher fields will enhance these interactions sufficiently to account for the discrepancies with the data.

\section{CONCLUSION}
We reported here the first single crystal synthesis of a breathing pyrochlore. Single crystal neutron scattering data using high quality single crystals of the breathing pyrochlore \byzo{} was presented for the first time. Our neutron scattering data show essentially  dispersionless bands within the instrument resolution. The single tetrahedron model can overall explain the observed dependence on wave-vector for the individual excitation bands of the system. Discrepancies between experimental data and the single tetrahedron model are observed in the diffuse scattering data at higher fields ($\mu_{0}H>$ 3T) and $T$ = 0.1 K, indicating that physics not included in the single tetrahedron model is responsible. Our powder diffraction data and pair distribution function analysis show that there is no observable disorder in the samples, which further support the suggestion that inter-tetrahedron interactions are the most likely explanation for the discrepancies between the experimental data and the  predictions of the  single tetrahedron model. Tuning the strength of the inter-tetrahedron interactions in breathing pyrochlores away from the isolated limit where $d'/d \gg 1$, by changing the breathing ratio, may be a path to realizing exotic states in breathing pyrochlores generally. The array of possibilities for such tuning, e.g. chemical or hydrostatic pressure~\cite{MirebeauNature,MirebeauPRL}, or by synthesis of new members of this family with different rare-earth ions (other than Yb$^{3+}$), presents many avenues for future investigations.

\label{sec:conclusion}

\begin{acknowledgments}
We are thankful to Stephen Kuhn for fruitful discussions and for his helps with this project in its early stages. M.J.P.G. acknowledges the Canada Research Chair (Tier I) program for support. S.H. acknowledges support provided by funding from William M. Fairbank Chair in Physics at Duke University, and from the Powe Junior Faculty Enhancement Award. This research used resources at the High Flux Isotope Reactor and Spallation Neutron Source, a DOE Office of Science User Facility operated by the Oak Ridge National Laboratory. We acknowledge the support of the National Institute of Standards and Technology, U.S. Department of Commerce, in providing the neutron research facilities used in this work. The identification of any commercial product or trade name does not imply endorsement or recommendation by the National Institute of Standards and Technology. A portion of this work was performed at the National High Magnetic Field Laboratory, which is supported by the National Science Foundation Cooperative Agreement No. DMR1157490 and DMR-1644779, the State of Florida and the U.S. Department of Energy. 
\end{acknowledgments}

\section{DATA AVAILABILITY}
The data that support the findings of this study are available from the corresponding authors upon request.

\section{AUTHOR CONTRIBUTIONS}
Research conceived by S.H.; Single crystal sample was grown and characterized by C.M., W.S., and S.H.; Thermodynamics and TDO measurements were performed by S.D., Z.S., R.B., D.G., and S.H.; Neutron scattering experiments were performed by S.D., W.S., N.P.B., M.F., A.P., J.L., and S.H.; Theoretical calculations were conducted by J.G.R and M.J.P.G.; All results were discussed and analyzed by S.D., Z.S, J.G.R, M.J.P.G, and S.H.; Manuscript written by S.D., J.G.R, M.J.P.G, and S.H.; All authors commented on the manuscript.

\section{COMPETING INTERESTS}
The authors declare no competing interests.

\bibliographystyle{apsrev4-2}
%\clearpage
\bibliography{main}

%apsrev4-2.bst 2019-01-14 (MD) hand-edited version of apsrev4-1.bst
%Control: key (0)
%Control: author (72) initials jnrlst
%Control: editor formatted (1) identically to author
%Control: production of article title (-1) disabled
%Control: page (0) single
%Control: year (1) truncated
%Control: production of eprint (0) enabled
\begin{thebibliography}{63}%
\makeatletter
\providecommand \@ifxundefined [1]{%
 \@ifx{#1\undefined}
}%
\providecommand \@ifnum [1]{%
 \ifnum #1\expandafter \@firstoftwo
 \else \expandafter \@secondoftwo
 \fi
}%
\providecommand \@ifx [1]{%
 \ifx #1\expandafter \@firstoftwo
 \else \expandafter \@secondoftwo
 \fi
}%
\providecommand \natexlab [1]{#1}%
\providecommand \enquote  [1]{``#1''}%
\providecommand \bibnamefont  [1]{#1}%
\providecommand \bibfnamefont [1]{#1}%
\providecommand \citenamefont [1]{#1}%
\providecommand \href@noop [0]{\@secondoftwo}%
\providecommand \href [0]{\begingroup \@sanitize@url \@href}%
\providecommand \@href[1]{\@@startlink{#1}\@@href}%
\providecommand \@@href[1]{\endgroup#1\@@endlink}%
\providecommand \@sanitize@url [0]{\catcode `\\12\catcode `\$12\catcode
  `\&12\catcode `\#12\catcode `\^12\catcode `\_12\catcode `\%12\relax}%
\providecommand \@@startlink[1]{}%
\providecommand \@@endlink[0]{}%
\providecommand \url  [0]{\begingroup\@sanitize@url \@url }%
\providecommand \@url [1]{\endgroup\@href {#1}{\urlprefix }}%
\providecommand \urlprefix  [0]{URL }%
\providecommand \Eprint [0]{\href }%
\providecommand \doibase [0]{https://doi.org/}%
\providecommand \selectlanguage [0]{\@gobble}%
\providecommand \bibinfo  [0]{\@secondoftwo}%
\providecommand \bibfield  [0]{\@secondoftwo}%
\providecommand \translation [1]{[#1]}%
\providecommand \BibitemOpen [0]{}%
\providecommand \bibitemStop [0]{}%
\providecommand \bibitemNoStop [0]{.\EOS\space}%
\providecommand \EOS [0]{\spacefactor3000\relax}%
\providecommand \BibitemShut  [1]{\csname bibitem#1\endcsname}%
\let\auto@bib@innerbib\@empty
%</preamble>
\bibitem [{\citenamefont {Balents}(2010)}]{Balents2010}%
  \BibitemOpen
  \bibfield  {author} {\bibinfo {author} {\bibfnamefont {L.}~\bibnamefont
  {Balents}},\ }\href {https://doi.org/10.1038/nature08917} {\bibfield
  {journal} {\bibinfo  {journal} {Nature}\ }\textbf {\bibinfo {volume} {464}},\
  \bibinfo {pages} {199} (\bibinfo {year} {2010})}\BibitemShut {NoStop}%
\bibitem [{\citenamefont {C.~Lacroix}(2011)}]{MilaMendelsLacroix}%
  \BibitemOpen
  \bibfield  {author} {\bibinfo {author} {\bibfnamefont {F.~M.}\ \bibnamefont
  {C.~Lacroix}, \bibfnamefont {P.~Mendels}},\ }\href
  {https://doi.org/10.1007/978-3-642-10589-0} {\emph {\bibinfo {title}
  {Introduction to Frustrated Magnetism}}}\ (\bibinfo {year}
  {2011})\BibitemShut {NoStop}%
\bibitem [{\citenamefont {Gardner}\ \emph {et~al.}(2010)\citenamefont
  {Gardner}, \citenamefont {Gingras},\ and\ \citenamefont
  {Greedan}}]{GardnerRMP2010}%
  \BibitemOpen
  \bibfield  {author} {\bibinfo {author} {\bibfnamefont {J.~S.}\ \bibnamefont
  {Gardner}}, \bibinfo {author} {\bibfnamefont {M.~J.~P.}\ \bibnamefont
  {Gingras}},\ and\ \bibinfo {author} {\bibfnamefont {J.~E.}\ \bibnamefont
  {Greedan}},\ }\href {https://doi.org/10.1103/RevModPhys.82.53} {\bibfield
  {journal} {\bibinfo  {journal} {Rev. Mod. Phys.}\ }\textbf {\bibinfo {volume}
  {82}},\ \bibinfo {pages} {53} (\bibinfo {year} {2010})}\BibitemShut {NoStop}%
\bibitem [{\citenamefont {Hallas}\ \emph {et~al.}(2018)\citenamefont {Hallas},
  \citenamefont {Gaudet},\ and\ \citenamefont {Gaulin}}]{HallasXYAnnuRevCMP}%
  \BibitemOpen
  \bibfield  {author} {\bibinfo {author} {\bibfnamefont {A.~M.}\ \bibnamefont
  {Hallas}}, \bibinfo {author} {\bibfnamefont {J.}~\bibnamefont {Gaudet}},\
  and\ \bibinfo {author} {\bibfnamefont {B.~D.}\ \bibnamefont {Gaulin}},\
  }\href {https://doi.org/10.1146/annurev-conmatphys-031016-025218} {\bibfield
  {journal} {\bibinfo  {journal} {Annual Review of Condensed Matter Physics}\
  }\textbf {\bibinfo {volume} {9}},\ \bibinfo {pages} {105} (\bibinfo {year}
  {2018})}\BibitemShut {NoStop}%
\bibitem [{\citenamefont {Rau}\ and\ \citenamefont
  {Gingras}(2019)}]{RauGingrasAnnuRevCMP}%
  \BibitemOpen
  \bibfield  {author} {\bibinfo {author} {\bibfnamefont {J.~G.}\ \bibnamefont
  {Rau}}\ and\ \bibinfo {author} {\bibfnamefont {M.~J.}\ \bibnamefont
  {Gingras}},\ }\href
  {https://doi.org/10.1146/annurev-conmatphys-022317-110520} {\bibfield
  {journal} {\bibinfo  {journal} {Annual Review of Condensed Matter Physics}\
  }\textbf {\bibinfo {volume} {10}},\ \bibinfo {pages} {357} (\bibinfo {year}
  {2019})}\BibitemShut {NoStop}%
\bibitem [{\citenamefont {Bramwell}\ and\ \citenamefont
  {Gingras}(2001)}]{Bramwell1495}%
  \BibitemOpen
  \bibfield  {author} {\bibinfo {author} {\bibfnamefont {S.~T.}\ \bibnamefont
  {Bramwell}}\ and\ \bibinfo {author} {\bibfnamefont {M.~J.~P.}\ \bibnamefont
  {Gingras}},\ }\href {https://doi.org/https://doi.org/10.1126/science.1064761}
  {\bibfield  {journal} {\bibinfo  {journal} {Science}\ }\textbf {\bibinfo
  {volume} {294}},\ \bibinfo {pages} {1495} (\bibinfo {year}
  {2001})}\BibitemShut {NoStop}%
\bibitem [{\citenamefont {Udagawa}\ and\ \citenamefont
  {Jaubert}(2021)}]{Springer-spin-ice}%
  \BibitemOpen
  \bibfield  {author} {\bibinfo {author} {\bibfnamefont {M.}~\bibnamefont
  {Udagawa}}\ and\ \bibinfo {author} {\bibfnamefont {L.}~\bibnamefont
  {Jaubert}},\ }\href@noop {} {\emph {\bibinfo {title} {Spin Ice}}}\ (\bibinfo
  {publisher} {Springer-Verlag, Berlin},\ \bibinfo {year} {2021})\BibitemShut
  {NoStop}%
\bibitem [{\citenamefont {Harris}\ \emph {et~al.}(1997)\citenamefont {Harris},
  \citenamefont {Bramwell}, \citenamefont {McMorrow}, \citenamefont {Zeiske},\
  and\ \citenamefont {Godfrey}}]{Harris1997}%
  \BibitemOpen
  \bibfield  {author} {\bibinfo {author} {\bibfnamefont {M.~J.}\ \bibnamefont
  {Harris}}, \bibinfo {author} {\bibfnamefont {S.~T.}\ \bibnamefont
  {Bramwell}}, \bibinfo {author} {\bibfnamefont {D.~F.}\ \bibnamefont
  {McMorrow}}, \bibinfo {author} {\bibfnamefont {T.}~\bibnamefont {Zeiske}},\
  and\ \bibinfo {author} {\bibfnamefont {K.~W.}\ \bibnamefont {Godfrey}},\
  }\href {https://doi.org/10.1103/PhysRevLett.79.2554} {\bibfield  {journal}
  {\bibinfo  {journal} {Phys. Rev. Lett.}\ }\textbf {\bibinfo {volume} {79}},\
  \bibinfo {pages} {2554} (\bibinfo {year} {1997})}\BibitemShut {NoStop}%
\bibitem [{\citenamefont {Bramwell}\ \emph {et~al.}(2001)\citenamefont
  {Bramwell}, \citenamefont {Harris}, \citenamefont {den Hertog}, \citenamefont
  {Gingras}, \citenamefont {Gardner}, \citenamefont {McMorrow}, \citenamefont
  {Wildes}, \citenamefont {Cornelius}, \citenamefont {Champion}, \citenamefont
  {Melko},\ and\ \citenamefont {Fennell}}]{PhysRevLett.87.047205}%
  \BibitemOpen
  \bibfield  {author} {\bibinfo {author} {\bibfnamefont {S.~T.}\ \bibnamefont
  {Bramwell}}, \bibinfo {author} {\bibfnamefont {M.~J.}\ \bibnamefont
  {Harris}}, \bibinfo {author} {\bibfnamefont {B.~C.}\ \bibnamefont {den
  Hertog}}, \bibinfo {author} {\bibfnamefont {M.~J.~P.}\ \bibnamefont
  {Gingras}}, \bibinfo {author} {\bibfnamefont {J.~S.}\ \bibnamefont
  {Gardner}}, \bibinfo {author} {\bibfnamefont {D.~F.}\ \bibnamefont
  {McMorrow}}, \bibinfo {author} {\bibfnamefont {A.~R.}\ \bibnamefont
  {Wildes}}, \bibinfo {author} {\bibfnamefont {A.~L.}\ \bibnamefont
  {Cornelius}}, \bibinfo {author} {\bibfnamefont {J.~D.~M.}\ \bibnamefont
  {Champion}}, \bibinfo {author} {\bibfnamefont {R.~G.}\ \bibnamefont
  {Melko}},\ and\ \bibinfo {author} {\bibfnamefont {T.}~\bibnamefont
  {Fennell}},\ }\href {https://doi.org/10.1103/PhysRevLett.87.047205}
  {\bibfield  {journal} {\bibinfo  {journal} {Phys. Rev. Lett.}\ }\textbf
  {\bibinfo {volume} {87}},\ \bibinfo {pages} {047205} (\bibinfo {year}
  {2001})}\BibitemShut {NoStop}%
\bibitem [{\citenamefont {Ramirez}\ \emph {et~al.}(1999)\citenamefont
  {Ramirez}, \citenamefont {Hayashi}, \citenamefont {Cava}, \citenamefont
  {Siddharthan},\ and\ \citenamefont {Shastry}}]{Ramirez1999}%
  \BibitemOpen
  \bibfield  {author} {\bibinfo {author} {\bibfnamefont {A.~P.}\ \bibnamefont
  {Ramirez}}, \bibinfo {author} {\bibfnamefont {A.}~\bibnamefont {Hayashi}},
  \bibinfo {author} {\bibfnamefont {R.~J.}\ \bibnamefont {Cava}}, \bibinfo
  {author} {\bibfnamefont {R.}~\bibnamefont {Siddharthan}},\ and\ \bibinfo
  {author} {\bibfnamefont {B.~S.}\ \bibnamefont {Shastry}},\ }\href
  {https://doi.org/10.1038/20619} {\bibfield  {journal} {\bibinfo  {journal}
  {Nature}\ }\textbf {\bibinfo {volume} {399}},\ \bibinfo {pages} {333}
  (\bibinfo {year} {1999})}\BibitemShut {NoStop}%
\bibitem [{\citenamefont {Gingras}\ and\ \citenamefont
  {McClarty}(2014)}]{Gingras_2014}%
  \BibitemOpen
  \bibfield  {author} {\bibinfo {author} {\bibfnamefont {M.~J.~P.}\
  \bibnamefont {Gingras}}\ and\ \bibinfo {author} {\bibfnamefont {P.~A.}\
  \bibnamefont {McClarty}},\ }\href
  {https://doi.org/10.1088/0034-4885/77/5/056501} {\bibfield  {journal}
  {\bibinfo  {journal} {Reports on Progress in Physics}\ }\textbf {\bibinfo
  {volume} {77}},\ \bibinfo {pages} {056501} (\bibinfo {year}
  {2014})}\BibitemShut {NoStop}%
\bibitem [{\citenamefont {Chang}\ \emph {et~al.}(2012)\citenamefont {Chang},
  \citenamefont {Onoda}, \citenamefont {Su}, \citenamefont {Kao}, \citenamefont
  {Tsuei}, \citenamefont {Yasui}, \citenamefont {Kakurai},\ and\ \citenamefont
  {Lees}}]{ChangHiggs2012}%
  \BibitemOpen
  \bibfield  {author} {\bibinfo {author} {\bibfnamefont {L.-J.}\ \bibnamefont
  {Chang}}, \bibinfo {author} {\bibfnamefont {S.}~\bibnamefont {Onoda}},
  \bibinfo {author} {\bibfnamefont {Y.}~\bibnamefont {Su}}, \bibinfo {author}
  {\bibfnamefont {Y.-J.}\ \bibnamefont {Kao}}, \bibinfo {author} {\bibfnamefont
  {K.-D.}\ \bibnamefont {Tsuei}}, \bibinfo {author} {\bibfnamefont
  {Y.}~\bibnamefont {Yasui}}, \bibinfo {author} {\bibfnamefont
  {K.}~\bibnamefont {Kakurai}},\ and\ \bibinfo {author} {\bibfnamefont {M.~R.}\
  \bibnamefont {Lees}},\ }\href
  {https://doi.org/https://doi.org/10.1038/ncomms1989} {\bibfield  {journal}
  {\bibinfo  {journal} {Nat. Commun.}\ }\textbf {\bibinfo {volume} {3}},\
  \bibinfo {pages} {992} (\bibinfo {year} {2012})}\BibitemShut {NoStop}%
\bibitem [{\citenamefont {Stewart}\ \emph {et~al.}(2004)\citenamefont
  {Stewart}, \citenamefont {Ehlers}, \citenamefont {Wills}, \citenamefont
  {Bramwell},\ and\ \citenamefont {Gardner}}]{StewartJPCM2004}%
  \BibitemOpen
  \bibfield  {author} {\bibinfo {author} {\bibfnamefont {J.}~\bibnamefont
  {Stewart}}, \bibinfo {author} {\bibfnamefont {G.}~\bibnamefont {Ehlers}},
  \bibinfo {author} {\bibfnamefont {A.~S.}\ \bibnamefont {Wills}}, \bibinfo
  {author} {\bibfnamefont {S.~T.}\ \bibnamefont {Bramwell}},\ and\ \bibinfo
  {author} {\bibfnamefont {J.~S.}\ \bibnamefont {Gardner}},\ }\href
  {https://doi.org/https://doi.org/10.1088/0034-4885/77/5/056501} {\bibfield
  {journal} {\bibinfo  {journal} {J. Phys. Condens. Matter}\ }\textbf {\bibinfo
  {volume} {16}},\ \bibinfo {pages} {L321} (\bibinfo {year}
  {2004})}\BibitemShut {NoStop}%
\bibitem [{\citenamefont {Javanparast}\ \emph {et~al.}(2015)\citenamefont
  {Javanparast}, \citenamefont {Hao}, \citenamefont {Enjalran},\ and\
  \citenamefont {Gingras}}]{Javanparast}%
  \BibitemOpen
  \bibfield  {author} {\bibinfo {author} {\bibfnamefont {B.}~\bibnamefont
  {Javanparast}}, \bibinfo {author} {\bibfnamefont {Z.}~\bibnamefont {Hao}},
  \bibinfo {author} {\bibfnamefont {M.}~\bibnamefont {Enjalran}},\ and\
  \bibinfo {author} {\bibfnamefont {M.~J.~P.}\ \bibnamefont {Gingras}},\ }\href
  {https://doi.org/10.1103/PhysRevLett.114.130601} {\bibfield  {journal}
  {\bibinfo  {journal} {Phys. Rev. Lett.}\ }\textbf {\bibinfo {volume} {114}},\
  \bibinfo {pages} {130601} (\bibinfo {year} {2015})}\BibitemShut {NoStop}%
\bibitem [{\citenamefont {Zhitomirsky}\ \emph {et~al.}(2012)\citenamefont
  {Zhitomirsky}, \citenamefont {Gvozdikova}, \citenamefont {Holdsworth},\ and\
  \citenamefont {Moessner}}]{Zhitomirsky2012}%
  \BibitemOpen
  \bibfield  {author} {\bibinfo {author} {\bibfnamefont {M.~E.}\ \bibnamefont
  {Zhitomirsky}}, \bibinfo {author} {\bibfnamefont {M.~V.}\ \bibnamefont
  {Gvozdikova}}, \bibinfo {author} {\bibfnamefont {P.~C.~W.}\ \bibnamefont
  {Holdsworth}},\ and\ \bibinfo {author} {\bibfnamefont {R.}~\bibnamefont
  {Moessner}},\ }\href {https://doi.org/10.1103/PhysRevLett.109.077204}
  {\bibfield  {journal} {\bibinfo  {journal} {Phys. Rev. Lett.}\ }\textbf
  {\bibinfo {volume} {109}},\ \bibinfo {pages} {077204} (\bibinfo {year}
  {2012})}\BibitemShut {NoStop}%
\bibitem [{\citenamefont {Savary}\ \emph {et~al.}(2012)\citenamefont {Savary},
  \citenamefont {Ross}, \citenamefont {Gaulin}, \citenamefont {Ruff},\ and\
  \citenamefont {Balents}}]{Savary_Er2ti2O7}%
  \BibitemOpen
  \bibfield  {author} {\bibinfo {author} {\bibfnamefont {L.}~\bibnamefont
  {Savary}}, \bibinfo {author} {\bibfnamefont {K.~A.}\ \bibnamefont {Ross}},
  \bibinfo {author} {\bibfnamefont {B.~D.}\ \bibnamefont {Gaulin}}, \bibinfo
  {author} {\bibfnamefont {J.~P.~C.}\ \bibnamefont {Ruff}},\ and\ \bibinfo
  {author} {\bibfnamefont {L.}~\bibnamefont {Balents}},\ }\href
  {https://doi.org/10.1103/PhysRevLett.109.167201} {\bibfield  {journal}
  {\bibinfo  {journal} {Phys. Rev. Lett.}\ }\textbf {\bibinfo {volume} {109}},\
  \bibinfo {pages} {167201} (\bibinfo {year} {2012})}\BibitemShut {NoStop}%
\bibitem [{\citenamefont {Rau}\ \emph {et~al.}(2016{\natexlab{a}})\citenamefont
  {Rau}, \citenamefont {Petit},\ and\ \citenamefont
  {Gingras}}]{PhysRevB.93.184408}%
  \BibitemOpen
  \bibfield  {author} {\bibinfo {author} {\bibfnamefont {J.~G.}\ \bibnamefont
  {Rau}}, \bibinfo {author} {\bibfnamefont {S.}~\bibnamefont {Petit}},\ and\
  \bibinfo {author} {\bibfnamefont {M.~J.~P.}\ \bibnamefont {Gingras}},\ }\href
  {https://doi.org/10.1103/PhysRevB.93.184408} {\bibfield  {journal} {\bibinfo
  {journal} {Phys. Rev. B}\ }\textbf {\bibinfo {volume} {93}},\ \bibinfo
  {pages} {184408} (\bibinfo {year} {2016}{\natexlab{a}})}\BibitemShut
  {NoStop}%
\bibitem [{\citenamefont {Oitmaa}\ \emph {et~al.}(2013)\citenamefont {Oitmaa},
  \citenamefont {Singh}, \citenamefont {Javanparast}, \citenamefont {Day},
  \citenamefont {Bagheri},\ and\ \citenamefont {Gingras}}]{Oitmaa}%
  \BibitemOpen
  \bibfield  {author} {\bibinfo {author} {\bibfnamefont {J.}~\bibnamefont
  {Oitmaa}}, \bibinfo {author} {\bibfnamefont {R.~R.~P.}\ \bibnamefont
  {Singh}}, \bibinfo {author} {\bibfnamefont {B.}~\bibnamefont {Javanparast}},
  \bibinfo {author} {\bibfnamefont {A.~G.~R.}\ \bibnamefont {Day}}, \bibinfo
  {author} {\bibfnamefont {B.~V.}\ \bibnamefont {Bagheri}},\ and\ \bibinfo
  {author} {\bibfnamefont {M.~J.~P.}\ \bibnamefont {Gingras}},\ }\href
  {https://doi.org/10.1103/PhysRevB.88.220404} {\bibfield  {journal} {\bibinfo
  {journal} {Phys. Rev. B}\ }\textbf {\bibinfo {volume} {88}},\ \bibinfo
  {pages} {220404} (\bibinfo {year} {2013})}\BibitemShut {NoStop}%
\bibitem [{\citenamefont {Gardner}\ \emph {et~al.}(1999)\citenamefont
  {Gardner}, \citenamefont {Dunsiger}, \citenamefont {Gaulin}, \citenamefont
  {Gingras}, \citenamefont {Greedan}, \citenamefont {Kiefl}, \citenamefont
  {Lumsden}, \citenamefont {MacFarlane}, \citenamefont {Raju}, \citenamefont
  {Sonier}, \citenamefont {Swainson},\ and\ \citenamefont
  {Tun}}]{PhysRevLett.82.1012}%
  \BibitemOpen
  \bibfield  {author} {\bibinfo {author} {\bibfnamefont {J.~S.}\ \bibnamefont
  {Gardner}}, \bibinfo {author} {\bibfnamefont {S.~R.}\ \bibnamefont
  {Dunsiger}}, \bibinfo {author} {\bibfnamefont {B.~D.}\ \bibnamefont
  {Gaulin}}, \bibinfo {author} {\bibfnamefont {M.~J.~P.}\ \bibnamefont
  {Gingras}}, \bibinfo {author} {\bibfnamefont {J.~E.}\ \bibnamefont
  {Greedan}}, \bibinfo {author} {\bibfnamefont {R.~F.}\ \bibnamefont {Kiefl}},
  \bibinfo {author} {\bibfnamefont {M.~D.}\ \bibnamefont {Lumsden}}, \bibinfo
  {author} {\bibfnamefont {W.~A.}\ \bibnamefont {MacFarlane}}, \bibinfo
  {author} {\bibfnamefont {N.~P.}\ \bibnamefont {Raju}}, \bibinfo {author}
  {\bibfnamefont {J.~E.}\ \bibnamefont {Sonier}}, \bibinfo {author}
  {\bibfnamefont {I.}~\bibnamefont {Swainson}},\ and\ \bibinfo {author}
  {\bibfnamefont {Z.}~\bibnamefont {Tun}},\ }\href
  {https://doi.org/10.1103/PhysRevLett.82.1012} {\bibfield  {journal} {\bibinfo
   {journal} {Phys. Rev. Lett.}\ }\textbf {\bibinfo {volume} {82}},\ \bibinfo
  {pages} {1012} (\bibinfo {year} {1999})}\BibitemShut {NoStop}%
\bibitem [{\citenamefont {Kimura}\ \emph {et~al.}(2013)\citenamefont {Kimura},
  \citenamefont {Nakatsuji}, \citenamefont {Wen}, \citenamefont {Broholm},
  \citenamefont {Stone}, \citenamefont {Nishibori},\ and\ \citenamefont
  {Sawa}}]{KimuraQuantum2013}%
  \BibitemOpen
  \bibfield  {author} {\bibinfo {author} {\bibfnamefont {K.}~\bibnamefont
  {Kimura}}, \bibinfo {author} {\bibfnamefont {S.}~\bibnamefont {Nakatsuji}},
  \bibinfo {author} {\bibfnamefont {J.~J.}\ \bibnamefont {Wen}}, \bibinfo
  {author} {\bibfnamefont {C.}~\bibnamefont {Broholm}}, \bibinfo {author}
  {\bibfnamefont {M.~B.}\ \bibnamefont {Stone}}, \bibinfo {author}
  {\bibfnamefont {E.}~\bibnamefont {Nishibori}},\ and\ \bibinfo {author}
  {\bibfnamefont {H.}~\bibnamefont {Sawa}},\ }\href
  {https://doi.org/10.1038/ncomms2914} {\bibfield  {journal} {\bibinfo
  {journal} {Nat. Commun.}\ }\textbf {\bibinfo {volume} {4}},\ \bibinfo {pages}
  {1934} (\bibinfo {year} {2013})}\BibitemShut {NoStop}%
\bibitem [{\citenamefont {Gao}\ \emph {et~al.}(2019)\citenamefont {Gao},
  \citenamefont {Chen}, \citenamefont {Tam}, \citenamefont {Huang},
  \citenamefont {Sasmal}, \citenamefont {Adroja}, \citenamefont {Ye},
  \citenamefont {Cao}, \citenamefont {Sala}, \citenamefont {Stone},
  \citenamefont {Baines}, \citenamefont {Verezhak}, \citenamefont {Hu},
  \citenamefont {Chung}, \citenamefont {Xu}, \citenamefont {Cheong},
  \citenamefont {Nallaiyan}, \citenamefont {Spagna}, \citenamefont {Maple},
  \citenamefont {Nevidomskyy}, \citenamefont {Morosan}, \citenamefont {Chen},\
  and\ \citenamefont {Dai}}]{GaoNatPhys2019}%
  \BibitemOpen
  \bibfield  {author} {\bibinfo {author} {\bibfnamefont {B.}~\bibnamefont
  {Gao}}, \bibinfo {author} {\bibfnamefont {T.}~\bibnamefont {Chen}}, \bibinfo
  {author} {\bibfnamefont {D.~W.}\ \bibnamefont {Tam}}, \bibinfo {author}
  {\bibfnamefont {C.-L.}\ \bibnamefont {Huang}}, \bibinfo {author}
  {\bibfnamefont {K.}~\bibnamefont {Sasmal}}, \bibinfo {author} {\bibfnamefont
  {D.~T.}\ \bibnamefont {Adroja}}, \bibinfo {author} {\bibfnamefont
  {F.}~\bibnamefont {Ye}}, \bibinfo {author} {\bibfnamefont {H.}~\bibnamefont
  {Cao}}, \bibinfo {author} {\bibfnamefont {G.}~\bibnamefont {Sala}}, \bibinfo
  {author} {\bibfnamefont {M.~B.}\ \bibnamefont {Stone}}, \bibinfo {author}
  {\bibfnamefont {C.}~\bibnamefont {Baines}}, \bibinfo {author} {\bibfnamefont
  {J.~A.~T.}\ \bibnamefont {Verezhak}}, \bibinfo {author} {\bibfnamefont
  {H.}~\bibnamefont {Hu}}, \bibinfo {author} {\bibfnamefont {J.-H.}\
  \bibnamefont {Chung}}, \bibinfo {author} {\bibfnamefont {X.}~\bibnamefont
  {Xu}}, \bibinfo {author} {\bibfnamefont {S.-W.}\ \bibnamefont {Cheong}},
  \bibinfo {author} {\bibfnamefont {M.}~\bibnamefont {Nallaiyan}}, \bibinfo
  {author} {\bibfnamefont {S.}~\bibnamefont {Spagna}}, \bibinfo {author}
  {\bibfnamefont {M.~B.}\ \bibnamefont {Maple}}, \bibinfo {author}
  {\bibfnamefont {A.~H.}\ \bibnamefont {Nevidomskyy}}, \bibinfo {author}
  {\bibfnamefont {E.}~\bibnamefont {Morosan}}, \bibinfo {author} {\bibfnamefont
  {G.}~\bibnamefont {Chen}},\ and\ \bibinfo {author} {\bibfnamefont
  {P.}~\bibnamefont {Dai}},\ }\href
  {https://doi.org/https://doi.org/10.1038/s41567-019-0577-6} {\bibfield
  {journal} {\bibinfo  {journal} {Nature Physics}\ }\textbf {\bibinfo {volume}
  {15}},\ \bibinfo {pages} {1052} (\bibinfo {year} {2019})}\BibitemShut
  {NoStop}%
\bibitem [{\citenamefont {Lee}\ \emph {et~al.}(2016)\citenamefont {Lee},
  \citenamefont {Do}, \citenamefont {Lee}, \citenamefont {Choi}, \citenamefont
  {Lee}, \citenamefont {Choi}, \citenamefont {Reyes}, \citenamefont {Kuhns},
  \citenamefont {Ozarowski},\ and\ \citenamefont {Choi}}]{LeePRB2016}%
  \BibitemOpen
  \bibfield  {author} {\bibinfo {author} {\bibfnamefont {S.}~\bibnamefont
  {Lee}}, \bibinfo {author} {\bibfnamefont {S.-H.}\ \bibnamefont {Do}},
  \bibinfo {author} {\bibfnamefont {W.-J.}\ \bibnamefont {Lee}}, \bibinfo
  {author} {\bibfnamefont {Y.~S.}\ \bibnamefont {Choi}}, \bibinfo {author}
  {\bibfnamefont {M.}~\bibnamefont {Lee}}, \bibinfo {author} {\bibfnamefont
  {E.~S.}\ \bibnamefont {Choi}}, \bibinfo {author} {\bibfnamefont {A.~P.}\
  \bibnamefont {Reyes}}, \bibinfo {author} {\bibfnamefont {P.~L.}\ \bibnamefont
  {Kuhns}}, \bibinfo {author} {\bibfnamefont {A.}~\bibnamefont {Ozarowski}},\
  and\ \bibinfo {author} {\bibfnamefont {K.-Y.}\ \bibnamefont {Choi}},\ }\href
  {https://doi.org/10.1103/PhysRevB.93.174402} {\bibfield  {journal} {\bibinfo
  {journal} {Phys. Rev. B}\ }\textbf {\bibinfo {volume} {93}},\ \bibinfo
  {pages} {174402} (\bibinfo {year} {2016})}\BibitemShut {NoStop}%
\bibitem [{\citenamefont {Okamoto}\ \emph {et~al.}(2013)\citenamefont
  {Okamoto}, \citenamefont {Nilsen}, \citenamefont {Attfield},\ and\
  \citenamefont {Hiroi}}]{Okamoto2013}%
  \BibitemOpen
  \bibfield  {author} {\bibinfo {author} {\bibfnamefont {Y.}~\bibnamefont
  {Okamoto}}, \bibinfo {author} {\bibfnamefont {G.~J.}\ \bibnamefont {Nilsen}},
  \bibinfo {author} {\bibfnamefont {J.~P.}\ \bibnamefont {Attfield}},\ and\
  \bibinfo {author} {\bibfnamefont {Z.}~\bibnamefont {Hiroi}},\ }\href
  {https://doi.org/10.1103/PhysRevLett.110.097203} {\bibfield  {journal}
  {\bibinfo  {journal} {Phys. Rev. Lett.}\ }\textbf {\bibinfo {volume} {110}},\
  \bibinfo {pages} {097203} (\bibinfo {year} {2013})}\BibitemShut {NoStop}%
\bibitem [{\citenamefont {Okamoto}\ \emph {et~al.}(2015)\citenamefont
  {Okamoto}, \citenamefont {Nilsen}, \citenamefont {Nakazono},\ and\
  \citenamefont {Hiroi}}]{OkamotoJPSJ2015}%
  \BibitemOpen
  \bibfield  {author} {\bibinfo {author} {\bibfnamefont {Y.}~\bibnamefont
  {Okamoto}}, \bibinfo {author} {\bibfnamefont {G.~J.}\ \bibnamefont {Nilsen}},
  \bibinfo {author} {\bibfnamefont {T.}~\bibnamefont {Nakazono}},\ and\
  \bibinfo {author} {\bibfnamefont {Z.}~\bibnamefont {Hiroi}},\ }\href
  {https://doi.org/10.7566/JPSJ.84.043707} {\bibfield  {journal} {\bibinfo
  {journal} {Journal of the Physical Society of Japan}\ }\textbf {\bibinfo
  {volume} {84}},\ \bibinfo {pages} {043707} (\bibinfo {year}
  {2015})}\BibitemShut {NoStop}%
\bibitem [{\citenamefont {Okamoto}\ \emph {et~al.}(2018)\citenamefont
  {Okamoto}, \citenamefont {Mori}, \citenamefont {Katayama}, \citenamefont
  {Miyake}, \citenamefont {Tokunaga}, \citenamefont {Matsuo}, \citenamefont
  {Kindo},\ and\ \citenamefont {Takenaka}}]{OkamotoJPSJ2018}%
  \BibitemOpen
  \bibfield  {author} {\bibinfo {author} {\bibfnamefont {Y.}~\bibnamefont
  {Okamoto}}, \bibinfo {author} {\bibfnamefont {M.}~\bibnamefont {Mori}},
  \bibinfo {author} {\bibfnamefont {N.}~\bibnamefont {Katayama}}, \bibinfo
  {author} {\bibfnamefont {A.}~\bibnamefont {Miyake}}, \bibinfo {author}
  {\bibfnamefont {M.}~\bibnamefont {Tokunaga}}, \bibinfo {author}
  {\bibfnamefont {A.}~\bibnamefont {Matsuo}}, \bibinfo {author} {\bibfnamefont
  {K.}~\bibnamefont {Kindo}},\ and\ \bibinfo {author} {\bibfnamefont
  {K.}~\bibnamefont {Takenaka}},\ }\href
  {https://doi.org/10.7566/JPSJ.87.034709} {\bibfield  {journal} {\bibinfo
  {journal} {Journal of the Physical Society of Japan}\ }\textbf {\bibinfo
  {volume} {87}},\ \bibinfo {pages} {034709} (\bibinfo {year}
  {2018})}\BibitemShut {NoStop}%
\bibitem [{\citenamefont {Pokharel}\ \emph {et~al.}(2018)\citenamefont
  {Pokharel}, \citenamefont {May}, \citenamefont {Parker}, \citenamefont
  {Calder}, \citenamefont {Ehlers}, \citenamefont {Huq}, \citenamefont
  {Kimber}, \citenamefont {Arachchige}, \citenamefont {Poudel}, \citenamefont
  {McGuire}, \citenamefont {Mandrus},\ and\ \citenamefont
  {Christianson}}]{PokharelPRB2016}%
  \BibitemOpen
  \bibfield  {author} {\bibinfo {author} {\bibfnamefont {G.}~\bibnamefont
  {Pokharel}}, \bibinfo {author} {\bibfnamefont {A.~F.}\ \bibnamefont {May}},
  \bibinfo {author} {\bibfnamefont {D.~S.}\ \bibnamefont {Parker}}, \bibinfo
  {author} {\bibfnamefont {S.}~\bibnamefont {Calder}}, \bibinfo {author}
  {\bibfnamefont {G.}~\bibnamefont {Ehlers}}, \bibinfo {author} {\bibfnamefont
  {A.}~\bibnamefont {Huq}}, \bibinfo {author} {\bibfnamefont {S.~A.~J.}\
  \bibnamefont {Kimber}}, \bibinfo {author} {\bibfnamefont {H.~S.}\
  \bibnamefont {Arachchige}}, \bibinfo {author} {\bibfnamefont
  {L.}~\bibnamefont {Poudel}}, \bibinfo {author} {\bibfnamefont {M.~A.}\
  \bibnamefont {McGuire}}, \bibinfo {author} {\bibfnamefont {D.}~\bibnamefont
  {Mandrus}},\ and\ \bibinfo {author} {\bibfnamefont {A.~D.}\ \bibnamefont
  {Christianson}},\ }\href {https://doi.org/10.1103/PhysRevB.97.134117}
  {\bibfield  {journal} {\bibinfo  {journal} {Phys. Rev. B}\ }\textbf {\bibinfo
  {volume} {97}},\ \bibinfo {pages} {134117} (\bibinfo {year}
  {2018})}\BibitemShut {NoStop}%
\bibitem [{\citenamefont {Pinch}\ \emph {et~al.}(1970)\citenamefont {Pinch},
  \citenamefont {Woods},\ and\ \citenamefont {Lopatin}}]{PINCH1970425}%
  \BibitemOpen
  \bibfield  {author} {\bibinfo {author} {\bibfnamefont {H.}~\bibnamefont
  {Pinch}}, \bibinfo {author} {\bibfnamefont {M.}~\bibnamefont {Woods}},\ and\
  \bibinfo {author} {\bibfnamefont {E.}~\bibnamefont {Lopatin}},\ }\href
  {https://doi.org/https://doi.org/10.1016/0025-5408(70)90081-4} {\bibfield
  {journal} {\bibinfo  {journal} {Materials Research Bulletin}\ }\textbf
  {\bibinfo {volume} {5}},\ \bibinfo {pages} {425 } (\bibinfo {year}
  {1970})}\BibitemShut {NoStop}%
\bibitem [{\citenamefont {Ghosh}\ \emph {et~al.}(2019)\citenamefont {Ghosh},
  \citenamefont {Iqbal}, \citenamefont {Müller}, \citenamefont {Ponnaganti},
  \citenamefont {Thomale}, \citenamefont {Narayanan}, \citenamefont {Reuther},
  \citenamefont {Gingras},\ and\ \citenamefont {Jeschke}}]{GhoshNPJQM2019}%
  \BibitemOpen
  \bibfield  {author} {\bibinfo {author} {\bibfnamefont {P.}~\bibnamefont
  {Ghosh}}, \bibinfo {author} {\bibfnamefont {Y.}~\bibnamefont {Iqbal}},
  \bibinfo {author} {\bibfnamefont {T.}~\bibnamefont {Müller}}, \bibinfo
  {author} {\bibfnamefont {R.~T.}\ \bibnamefont {Ponnaganti}}, \bibinfo
  {author} {\bibfnamefont {R.}~\bibnamefont {Thomale}}, \bibinfo {author}
  {\bibfnamefont {R.}~\bibnamefont {Narayanan}}, \bibinfo {author}
  {\bibfnamefont {J.}~\bibnamefont {Reuther}}, \bibinfo {author} {\bibfnamefont
  {M.~J.~P.}\ \bibnamefont {Gingras}},\ and\ \bibinfo {author} {\bibfnamefont
  {H.~O.}\ \bibnamefont {Jeschke}},\ }\href
  {https://doi.org/10.1038/s41535-019-0202-z} {\bibfield  {journal} {\bibinfo
  {journal} {npj Quantum Materials}\ }\textbf {\bibinfo {volume} {4}},\
  \bibinfo {pages} {63} (\bibinfo {year} {2019})}\BibitemShut {NoStop}%
\bibitem [{\citenamefont {Kimura}\ \emph {et~al.}(2014)\citenamefont {Kimura},
  \citenamefont {Nakatsuji},\ and\ \citenamefont {Kimura}}]{KimuraPRB2014}%
  \BibitemOpen
  \bibfield  {author} {\bibinfo {author} {\bibfnamefont {K.}~\bibnamefont
  {Kimura}}, \bibinfo {author} {\bibfnamefont {S.}~\bibnamefont {Nakatsuji}},\
  and\ \bibinfo {author} {\bibfnamefont {T.}~\bibnamefont {Kimura}},\ }\href
  {https://doi.org/10.1103/PhysRevB.90.060414} {\bibfield  {journal} {\bibinfo
  {journal} {Phys. Rev. B}\ }\textbf {\bibinfo {volume} {90}},\ \bibinfo
  {pages} {060414} (\bibinfo {year} {2014})}\BibitemShut {NoStop}%
\bibitem [{\citenamefont {Rau}\ \emph {et~al.}(2016{\natexlab{b}})\citenamefont
  {Rau}, \citenamefont {Wu}, \citenamefont {May}, \citenamefont {Poudel},
  \citenamefont {Winn}, \citenamefont {Garlea}, \citenamefont {Huq},
  \citenamefont {Whitfield}, \citenamefont {Taylor}, \citenamefont {Lumsden},
  \citenamefont {Gingras},\ and\ \citenamefont {Christianson}}]{RauPRL2016}%
  \BibitemOpen
  \bibfield  {author} {\bibinfo {author} {\bibfnamefont {J.~G.}\ \bibnamefont
  {Rau}}, \bibinfo {author} {\bibfnamefont {L.~S.}\ \bibnamefont {Wu}},
  \bibinfo {author} {\bibfnamefont {A.~F.}\ \bibnamefont {May}}, \bibinfo
  {author} {\bibfnamefont {L.}~\bibnamefont {Poudel}}, \bibinfo {author}
  {\bibfnamefont {B.}~\bibnamefont {Winn}}, \bibinfo {author} {\bibfnamefont
  {V.~O.}\ \bibnamefont {Garlea}}, \bibinfo {author} {\bibfnamefont
  {A.}~\bibnamefont {Huq}}, \bibinfo {author} {\bibfnamefont {P.}~\bibnamefont
  {Whitfield}}, \bibinfo {author} {\bibfnamefont {A.~E.}\ \bibnamefont
  {Taylor}}, \bibinfo {author} {\bibfnamefont {M.~D.}\ \bibnamefont {Lumsden}},
  \bibinfo {author} {\bibfnamefont {M.~J.~P.}\ \bibnamefont {Gingras}},\ and\
  \bibinfo {author} {\bibfnamefont {A.~D.}\ \bibnamefont {Christianson}},\
  }\href {https://doi.org/10.1103/PhysRevLett.116.257204} {\bibfield  {journal}
  {\bibinfo  {journal} {Phys. Rev. Lett.}\ }\textbf {\bibinfo {volume} {116}},\
  \bibinfo {pages} {257204} (\bibinfo {year} {2016}{\natexlab{b}})}\BibitemShut
  {NoStop}%
\bibitem [{\citenamefont {Haku}\ \emph
  {et~al.}(2016{\natexlab{a}})\citenamefont {Haku}, \citenamefont {Kimura},
  \citenamefont {Matsumoto}, \citenamefont {Soda}, \citenamefont {Sera},
  \citenamefont {Yu}, \citenamefont {Mole}, \citenamefont {Takeuchi},
  \citenamefont {Nakatsuji}, \citenamefont {Kono}, \citenamefont {Sakakibara},
  \citenamefont {Chang},\ and\ \citenamefont {Masuda}}]{HakuPRB2016}%
  \BibitemOpen
  \bibfield  {author} {\bibinfo {author} {\bibfnamefont {T.}~\bibnamefont
  {Haku}}, \bibinfo {author} {\bibfnamefont {K.}~\bibnamefont {Kimura}},
  \bibinfo {author} {\bibfnamefont {Y.}~\bibnamefont {Matsumoto}}, \bibinfo
  {author} {\bibfnamefont {M.}~\bibnamefont {Soda}}, \bibinfo {author}
  {\bibfnamefont {M.}~\bibnamefont {Sera}}, \bibinfo {author} {\bibfnamefont
  {D.}~\bibnamefont {Yu}}, \bibinfo {author} {\bibfnamefont {R.~A.}\
  \bibnamefont {Mole}}, \bibinfo {author} {\bibfnamefont {T.}~\bibnamefont
  {Takeuchi}}, \bibinfo {author} {\bibfnamefont {S.}~\bibnamefont {Nakatsuji}},
  \bibinfo {author} {\bibfnamefont {Y.}~\bibnamefont {Kono}}, \bibinfo {author}
  {\bibfnamefont {T.}~\bibnamefont {Sakakibara}}, \bibinfo {author}
  {\bibfnamefont {L.-J.}\ \bibnamefont {Chang}},\ and\ \bibinfo {author}
  {\bibfnamefont {T.}~\bibnamefont {Masuda}},\ }\href
  {https://doi.org/10.1103/PhysRevB.93.220407} {\bibfield  {journal} {\bibinfo
  {journal} {Phys. Rev. B}\ }\textbf {\bibinfo {volume} {93}},\ \bibinfo
  {pages} {220407} (\bibinfo {year} {2016}{\natexlab{a}})}\BibitemShut
  {NoStop}%
\bibitem [{\citenamefont {Haku}\ \emph
  {et~al.}(2016{\natexlab{b}})\citenamefont {Haku}, \citenamefont {Soda},
  \citenamefont {Sera}, \citenamefont {Kimura}, \citenamefont {Itoh},
  \citenamefont {Yokoo},\ and\ \citenamefont {Masuda}}]{HakuJPSJ2016}%
  \BibitemOpen
  \bibfield  {author} {\bibinfo {author} {\bibfnamefont {T.}~\bibnamefont
  {Haku}}, \bibinfo {author} {\bibfnamefont {M.}~\bibnamefont {Soda}}, \bibinfo
  {author} {\bibfnamefont {M.}~\bibnamefont {Sera}}, \bibinfo {author}
  {\bibfnamefont {K.}~\bibnamefont {Kimura}}, \bibinfo {author} {\bibfnamefont
  {S.}~\bibnamefont {Itoh}}, \bibinfo {author} {\bibfnamefont {T.}~\bibnamefont
  {Yokoo}},\ and\ \bibinfo {author} {\bibfnamefont {T.}~\bibnamefont
  {Masuda}},\ }\href {https://doi.org/10.7566/JPSJ.85.034721} {\bibfield
  {journal} {\bibinfo  {journal} {Journal of the Physical Society of Japan}\
  }\textbf {\bibinfo {volume} {85}},\ \bibinfo {pages} {034721} (\bibinfo
  {year} {2016}{\natexlab{b}})}\BibitemShut {NoStop}%
\bibitem [{\citenamefont {Park}\ \emph {et~al.}(2016)\citenamefont {Park},
  \citenamefont {Do}, \citenamefont {Choi}, \citenamefont {Kang}, \citenamefont
  {Jang}, \citenamefont {Schmidt}, \citenamefont {Brando}, \citenamefont {Kim},
  \citenamefont {Kim}, \citenamefont {Butch}, \citenamefont {Lee},
  \citenamefont {Park},\ and\ \citenamefont {Ji}}]{Park2016}%
  \BibitemOpen
  \bibfield  {author} {\bibinfo {author} {\bibfnamefont {S.-Y.}\ \bibnamefont
  {Park}}, \bibinfo {author} {\bibfnamefont {S.-H.}\ \bibnamefont {Do}},
  \bibinfo {author} {\bibfnamefont {K.-Y.}\ \bibnamefont {Choi}}, \bibinfo
  {author} {\bibfnamefont {J.-H.}\ \bibnamefont {Kang}}, \bibinfo {author}
  {\bibfnamefont {D.}~\bibnamefont {Jang}}, \bibinfo {author} {\bibfnamefont
  {B.}~\bibnamefont {Schmidt}}, \bibinfo {author} {\bibfnamefont
  {M.}~\bibnamefont {Brando}}, \bibinfo {author} {\bibfnamefont {B.-H.}\
  \bibnamefont {Kim}}, \bibinfo {author} {\bibfnamefont {D.-H.}\ \bibnamefont
  {Kim}}, \bibinfo {author} {\bibfnamefont {N.~P.}\ \bibnamefont {Butch}},
  \bibinfo {author} {\bibfnamefont {S.}~\bibnamefont {Lee}}, \bibinfo {author}
  {\bibfnamefont {J.-H.}\ \bibnamefont {Park}},\ and\ \bibinfo {author}
  {\bibfnamefont {S.}~\bibnamefont {Ji}},\ }\href
  {https://doi.org/10.1038/ncomms12912} {\bibfield  {journal} {\bibinfo
  {journal} {Nature Communications}\ }\textbf {\bibinfo {volume} {7}},\
  \bibinfo {pages} {12912} (\bibinfo {year} {2016})}\BibitemShut {NoStop}%
\bibitem [{\citenamefont {Rau}\ and\ \citenamefont
  {Gingras}(2018)}]{RauPRB2018}%
  \BibitemOpen
  \bibfield  {author} {\bibinfo {author} {\bibfnamefont {J.~G.}\ \bibnamefont
  {Rau}}\ and\ \bibinfo {author} {\bibfnamefont {M.~J.~P.}\ \bibnamefont
  {Gingras}},\ }\href {https://doi.org/10.1103/PhysRevB.98.054408} {\bibfield
  {journal} {\bibinfo  {journal} {Phys. Rev. B}\ }\textbf {\bibinfo {volume}
  {98}},\ \bibinfo {pages} {054408} (\bibinfo {year} {2018})}\BibitemShut
  {NoStop}%
\bibitem [{\citenamefont {Rau}\ \emph {et~al.}(2018)\citenamefont {Rau},
  \citenamefont {Wu}, \citenamefont {May}, \citenamefont {Taylor},
  \citenamefont {Liu}, \citenamefont {Higgins}, \citenamefont {Butch},
  \citenamefont {Ross}, \citenamefont {Nair}, \citenamefont {Lumsden},
  \citenamefont {Gingras},\ and\ \citenamefont {Christianson}}]{RauJPCM_2018}%
  \BibitemOpen
  \bibfield  {author} {\bibinfo {author} {\bibfnamefont {J.~G.}\ \bibnamefont
  {Rau}}, \bibinfo {author} {\bibfnamefont {L.~S.}\ \bibnamefont {Wu}},
  \bibinfo {author} {\bibfnamefont {A.~F.}\ \bibnamefont {May}}, \bibinfo
  {author} {\bibfnamefont {A.~E.}\ \bibnamefont {Taylor}}, \bibinfo {author}
  {\bibfnamefont {I.-L.}\ \bibnamefont {Liu}}, \bibinfo {author} {\bibfnamefont
  {J.}~\bibnamefont {Higgins}}, \bibinfo {author} {\bibfnamefont {N.~P.}\
  \bibnamefont {Butch}}, \bibinfo {author} {\bibfnamefont {K.~A.}\ \bibnamefont
  {Ross}}, \bibinfo {author} {\bibfnamefont {H.~S.}\ \bibnamefont {Nair}},
  \bibinfo {author} {\bibfnamefont {M.~D.}\ \bibnamefont {Lumsden}}, \bibinfo
  {author} {\bibfnamefont {M.~J.~P.}\ \bibnamefont {Gingras}},\ and\ \bibinfo
  {author} {\bibfnamefont {A.~D.}\ \bibnamefont {Christianson}},\ }\href
  {https://doi.org/10.1088/1361-648x/aae45a} {\bibfield  {journal} {\bibinfo
  {journal} {Journal of Physics: Condensed Matter}\ }\textbf {\bibinfo {volume}
  {30}},\ \bibinfo {pages} {455801} (\bibinfo {year} {2018})}\BibitemShut
  {NoStop}%
\bibitem [{\citenamefont {Higo}\ \emph {et~al.}(2017)\citenamefont {Higo},
  \citenamefont {Iritani}, \citenamefont {Halim}, \citenamefont {Higemoto},
  \citenamefont {Ito}, \citenamefont {Kuga}, \citenamefont {Kimura},\ and\
  \citenamefont {Nakatsuji}}]{spinel1}%
  \BibitemOpen
  \bibfield  {author} {\bibinfo {author} {\bibfnamefont {T.}~\bibnamefont
  {Higo}}, \bibinfo {author} {\bibfnamefont {K.}~\bibnamefont {Iritani}},
  \bibinfo {author} {\bibfnamefont {M.}~\bibnamefont {Halim}}, \bibinfo
  {author} {\bibfnamefont {W.}~\bibnamefont {Higemoto}}, \bibinfo {author}
  {\bibfnamefont {T.~U.}\ \bibnamefont {Ito}}, \bibinfo {author} {\bibfnamefont
  {K.}~\bibnamefont {Kuga}}, \bibinfo {author} {\bibfnamefont {K.}~\bibnamefont
  {Kimura}},\ and\ \bibinfo {author} {\bibfnamefont {S.}~\bibnamefont
  {Nakatsuji}},\ }\href {https://doi.org/10.1103/PhysRevB.95.174443} {\bibfield
   {journal} {\bibinfo  {journal} {Phys. Rev. B}\ }\textbf {\bibinfo {volume}
  {95}},\ \bibinfo {pages} {174443} (\bibinfo {year} {2017})}\BibitemShut
  {NoStop}%
\bibitem [{\citenamefont {Dalmas~de R\'eotier}\ \emph
  {et~al.}(2017)\citenamefont {Dalmas~de R\'eotier}, \citenamefont {Marin},
  \citenamefont {Yaouanc}, \citenamefont {Ritter}, \citenamefont {Maisuradze},
  \citenamefont {Roessli}, \citenamefont {Bertin}, \citenamefont {Baker},\ and\
  \citenamefont {Amato}}]{spinel2}%
  \BibitemOpen
  \bibfield  {author} {\bibinfo {author} {\bibfnamefont {P.}~\bibnamefont
  {Dalmas~de R\'eotier}}, \bibinfo {author} {\bibfnamefont {C.}~\bibnamefont
  {Marin}}, \bibinfo {author} {\bibfnamefont {A.}~\bibnamefont {Yaouanc}},
  \bibinfo {author} {\bibfnamefont {C.}~\bibnamefont {Ritter}}, \bibinfo
  {author} {\bibfnamefont {A.}~\bibnamefont {Maisuradze}}, \bibinfo {author}
  {\bibfnamefont {B.}~\bibnamefont {Roessli}}, \bibinfo {author} {\bibfnamefont
  {A.}~\bibnamefont {Bertin}}, \bibinfo {author} {\bibfnamefont {P.~J.}\
  \bibnamefont {Baker}},\ and\ \bibinfo {author} {\bibfnamefont
  {A.}~\bibnamefont {Amato}},\ }\href
  {https://doi.org/10.1103/PhysRevB.96.134403} {\bibfield  {journal} {\bibinfo
  {journal} {Phys. Rev. B}\ }\textbf {\bibinfo {volume} {96}},\ \bibinfo
  {pages} {134403} (\bibinfo {year} {2017})}\BibitemShut {NoStop}%
\bibitem [{\citenamefont {Guratinder}\ \emph {et~al.}(2019)\citenamefont
  {Guratinder}, \citenamefont {Rau}, \citenamefont {Tsurkan}, \citenamefont
  {Ritter}, \citenamefont {Embs}, \citenamefont {Fennell}, \citenamefont
  {Walker}, \citenamefont {Medarde}, \citenamefont {Shang}, \citenamefont
  {Cervellino}, \citenamefont {R\"uegg},\ and\ \citenamefont
  {Zaharko}}]{spinel3}%
  \BibitemOpen
  \bibfield  {author} {\bibinfo {author} {\bibfnamefont {K.}~\bibnamefont
  {Guratinder}}, \bibinfo {author} {\bibfnamefont {J.~G.}\ \bibnamefont {Rau}},
  \bibinfo {author} {\bibfnamefont {V.}~\bibnamefont {Tsurkan}}, \bibinfo
  {author} {\bibfnamefont {C.}~\bibnamefont {Ritter}}, \bibinfo {author}
  {\bibfnamefont {J.}~\bibnamefont {Embs}}, \bibinfo {author} {\bibfnamefont
  {T.}~\bibnamefont {Fennell}}, \bibinfo {author} {\bibfnamefont {H.~C.}\
  \bibnamefont {Walker}}, \bibinfo {author} {\bibfnamefont {M.}~\bibnamefont
  {Medarde}}, \bibinfo {author} {\bibfnamefont {T.}~\bibnamefont {Shang}},
  \bibinfo {author} {\bibfnamefont {A.}~\bibnamefont {Cervellino}}, \bibinfo
  {author} {\bibfnamefont {C.}~\bibnamefont {R\"uegg}},\ and\ \bibinfo {author}
  {\bibfnamefont {O.}~\bibnamefont {Zaharko}},\ }\href
  {https://doi.org/10.1103/PhysRevB.100.094420} {\bibfield  {journal} {\bibinfo
   {journal} {Phys. Rev. B}\ }\textbf {\bibinfo {volume} {100}},\ \bibinfo
  {pages} {094420} (\bibinfo {year} {2019})}\BibitemShut {NoStop}%
\bibitem [{\citenamefont {Xing}\ \emph {et~al.}(2020)\citenamefont {Xing},
  \citenamefont {Feng}, \citenamefont {Liu}, \citenamefont {Emmanouilidou},
  \citenamefont {Hu}, \citenamefont {Liu}, \citenamefont {Graf}, \citenamefont
  {Ramirez}, \citenamefont {Chen}, \citenamefont {Cao},\ and\ \citenamefont
  {Ni}}]{ybcl31}%
  \BibitemOpen
  \bibfield  {author} {\bibinfo {author} {\bibfnamefont {J.}~\bibnamefont
  {Xing}}, \bibinfo {author} {\bibfnamefont {E.}~\bibnamefont {Feng}}, \bibinfo
  {author} {\bibfnamefont {Y.}~\bibnamefont {Liu}}, \bibinfo {author}
  {\bibfnamefont {E.}~\bibnamefont {Emmanouilidou}}, \bibinfo {author}
  {\bibfnamefont {C.}~\bibnamefont {Hu}}, \bibinfo {author} {\bibfnamefont
  {J.}~\bibnamefont {Liu}}, \bibinfo {author} {\bibfnamefont {D.}~\bibnamefont
  {Graf}}, \bibinfo {author} {\bibfnamefont {A.~P.}\ \bibnamefont {Ramirez}},
  \bibinfo {author} {\bibfnamefont {G.}~\bibnamefont {Chen}}, \bibinfo {author}
  {\bibfnamefont {H.}~\bibnamefont {Cao}},\ and\ \bibinfo {author}
  {\bibfnamefont {N.}~\bibnamefont {Ni}},\ }\href
  {https://doi.org/10.1103/PhysRevB.102.014427} {\bibfield  {journal} {\bibinfo
   {journal} {Phys. Rev. B}\ }\textbf {\bibinfo {volume} {102}},\ \bibinfo
  {pages} {014427} (\bibinfo {year} {2020})}\BibitemShut {NoStop}%
\bibitem [{\citenamefont {Sala}\ \emph {et~al.}(2019)\citenamefont {Sala},
  \citenamefont {Stone}, \citenamefont {Rai}, \citenamefont {May},
  \citenamefont {Parker}, \citenamefont {Hal\'asz}, \citenamefont {Cheng},
  \citenamefont {Ehlers}, \citenamefont {Garlea}, \citenamefont {Zhang},
  \citenamefont {Lumsden},\ and\ \citenamefont {Christianson}}]{ybcl32}%
  \BibitemOpen
  \bibfield  {author} {\bibinfo {author} {\bibfnamefont {G.}~\bibnamefont
  {Sala}}, \bibinfo {author} {\bibfnamefont {M.~B.}\ \bibnamefont {Stone}},
  \bibinfo {author} {\bibfnamefont {B.~K.}\ \bibnamefont {Rai}}, \bibinfo
  {author} {\bibfnamefont {A.~F.}\ \bibnamefont {May}}, \bibinfo {author}
  {\bibfnamefont {D.~S.}\ \bibnamefont {Parker}}, \bibinfo {author}
  {\bibfnamefont {G.~B.}\ \bibnamefont {Hal\'asz}}, \bibinfo {author}
  {\bibfnamefont {Y.~Q.}\ \bibnamefont {Cheng}}, \bibinfo {author}
  {\bibfnamefont {G.}~\bibnamefont {Ehlers}}, \bibinfo {author} {\bibfnamefont
  {V.~O.}\ \bibnamefont {Garlea}}, \bibinfo {author} {\bibfnamefont
  {Q.}~\bibnamefont {Zhang}}, \bibinfo {author} {\bibfnamefont {M.~D.}\
  \bibnamefont {Lumsden}},\ and\ \bibinfo {author} {\bibfnamefont {A.~D.}\
  \bibnamefont {Christianson}},\ }\href
  {https://doi.org/10.1103/PhysRevB.100.180406} {\bibfield  {journal} {\bibinfo
   {journal} {Phys. Rev. B}\ }\textbf {\bibinfo {volume} {100}},\ \bibinfo
  {pages} {180406} (\bibinfo {year} {2019})}\BibitemShut {NoStop}%
\bibitem [{\citenamefont {Sala}\ \emph {et~al.}(2020)\citenamefont {Sala},
  \citenamefont {Stone}, \citenamefont {Rai}, \citenamefont {May},
  \citenamefont {Laurell}, \citenamefont {Garlea}, \citenamefont {Butch},
  \citenamefont {Lumsden}, \citenamefont {Ehlers}, \citenamefont {Pokharel}
  \emph {et~al.}}]{ybcl33}%
  \BibitemOpen
  \bibfield  {author} {\bibinfo {author} {\bibfnamefont {G.}~\bibnamefont
  {Sala}}, \bibinfo {author} {\bibfnamefont {M.}~\bibnamefont {Stone}},
  \bibinfo {author} {\bibfnamefont {B.~K.}\ \bibnamefont {Rai}}, \bibinfo
  {author} {\bibfnamefont {A.}~\bibnamefont {May}}, \bibinfo {author}
  {\bibfnamefont {P.}~\bibnamefont {Laurell}}, \bibinfo {author} {\bibfnamefont
  {V.}~\bibnamefont {Garlea}}, \bibinfo {author} {\bibfnamefont
  {N.}~\bibnamefont {Butch}}, \bibinfo {author} {\bibfnamefont
  {M.}~\bibnamefont {Lumsden}}, \bibinfo {author} {\bibfnamefont
  {G.}~\bibnamefont {Ehlers}}, \bibinfo {author} {\bibfnamefont
  {G.}~\bibnamefont {Pokharel}}, \emph {et~al.},\ }\href@noop {} {\bibfield
  {journal} {\bibinfo  {journal} {arXiv preprint arXiv:2003.01754}\ } (\bibinfo
  {year} {2020})}\BibitemShut {NoStop}%
\bibitem [{\citenamefont {Li}\ \emph {et~al.}(2016)\citenamefont {Li},
  \citenamefont {Li}, \citenamefont {Kim}, \citenamefont {Balents},
  \citenamefont {Yu},\ and\ \citenamefont {Chen}}]{Li2016}%
  \BibitemOpen
  \bibfield  {author} {\bibinfo {author} {\bibfnamefont {F.-Y.}\ \bibnamefont
  {Li}}, \bibinfo {author} {\bibfnamefont {Y.-D.}\ \bibnamefont {Li}}, \bibinfo
  {author} {\bibfnamefont {Y.~B.}\ \bibnamefont {Kim}}, \bibinfo {author}
  {\bibfnamefont {L.}~\bibnamefont {Balents}}, \bibinfo {author} {\bibfnamefont
  {Y.}~\bibnamefont {Yu}},\ and\ \bibinfo {author} {\bibfnamefont
  {G.}~\bibnamefont {Chen}},\ }\href {https://doi.org/10.1038/ncomms12691}
  {\bibfield  {journal} {\bibinfo  {journal} {Nature Communications}\ }\textbf
  {\bibinfo {volume} {7}},\ \bibinfo {pages} {12691} (\bibinfo {year}
  {2016})}\BibitemShut {NoStop}%
\bibitem [{\citenamefont {Savary}\ \emph {et~al.}(2016)\citenamefont {Savary},
  \citenamefont {Wang}, \citenamefont {Kee}, \citenamefont {Kim}, \citenamefont
  {Yu},\ and\ \citenamefont {Chen}}]{SavaryPRB2016}%
  \BibitemOpen
  \bibfield  {author} {\bibinfo {author} {\bibfnamefont {L.}~\bibnamefont
  {Savary}}, \bibinfo {author} {\bibfnamefont {X.}~\bibnamefont {Wang}},
  \bibinfo {author} {\bibfnamefont {H.-Y.}\ \bibnamefont {Kee}}, \bibinfo
  {author} {\bibfnamefont {Y.~B.}\ \bibnamefont {Kim}}, \bibinfo {author}
  {\bibfnamefont {Y.}~\bibnamefont {Yu}},\ and\ \bibinfo {author}
  {\bibfnamefont {G.}~\bibnamefont {Chen}},\ }\href
  {https://doi.org/10.1103/PhysRevB.94.075146} {\bibfield  {journal} {\bibinfo
  {journal} {Phys. Rev. B}\ }\textbf {\bibinfo {volume} {94}},\ \bibinfo
  {pages} {075146} (\bibinfo {year} {2016})}\BibitemShut {NoStop}%
\bibitem [{\citenamefont {Yan}\ \emph {et~al.}(2020)\citenamefont {Yan},
  \citenamefont {Benton}, \citenamefont {Jaubert},\ and\ \citenamefont
  {Shannon}}]{YanPRL2020}%
  \BibitemOpen
  \bibfield  {author} {\bibinfo {author} {\bibfnamefont {H.}~\bibnamefont
  {Yan}}, \bibinfo {author} {\bibfnamefont {O.}~\bibnamefont {Benton}},
  \bibinfo {author} {\bibfnamefont {L.~D.~C.}\ \bibnamefont {Jaubert}},\ and\
  \bibinfo {author} {\bibfnamefont {N.}~\bibnamefont {Shannon}},\ }\href
  {https://doi.org/10.1103/PhysRevLett.124.127203} {\bibfield  {journal}
  {\bibinfo  {journal} {Phys. Rev. Lett.}\ }\textbf {\bibinfo {volume} {124}},\
  \bibinfo {pages} {127203} (\bibinfo {year} {2020})}\BibitemShut {NoStop}%
\bibitem [{\citenamefont {Toby}(2006)}]{R_Factors}%
  \BibitemOpen
  \bibfield  {author} {\bibinfo {author} {\bibfnamefont {B.~H.}\ \bibnamefont
  {Toby}},\ }\href {https://doi.org/10.1154/1.2179804} {\bibfield  {journal}
  {\bibinfo  {journal} {Powder Diffraction}\ }\textbf {\bibinfo {volume}
  {21}},\ \bibinfo {pages} {67–70} (\bibinfo {year} {2006})}\BibitemShut
  {NoStop}%
\bibitem [{\citenamefont {Farrow}\ \emph {et~al.}(2007)\citenamefont {Farrow},
  \citenamefont {Juhas}, \citenamefont {Liu}, \citenamefont {Bryndin},
  \citenamefont {Bo{\v{z}}in}, \citenamefont {Bloch}, \citenamefont {Proffen},\
  and\ \citenamefont {Billinge}}]{Pdfgui}%
  \BibitemOpen
  \bibfield  {author} {\bibinfo {author} {\bibfnamefont {C.~L.}\ \bibnamefont
  {Farrow}}, \bibinfo {author} {\bibfnamefont {P.}~\bibnamefont {Juhas}},
  \bibinfo {author} {\bibfnamefont {J.~W.}\ \bibnamefont {Liu}}, \bibinfo
  {author} {\bibfnamefont {D.}~\bibnamefont {Bryndin}}, \bibinfo {author}
  {\bibfnamefont {E.~S.}\ \bibnamefont {Bo{\v{z}}in}}, \bibinfo {author}
  {\bibfnamefont {J.}~\bibnamefont {Bloch}}, \bibinfo {author} {\bibfnamefont
  {T.}~\bibnamefont {Proffen}},\ and\ \bibinfo {author} {\bibfnamefont
  {S.~J.~L.}\ \bibnamefont {Billinge}},\ }\href
  {https://doi.org/10.1088/0953-8984/19/33/335219} {\bibfield  {journal}
  {\bibinfo  {journal} {Journal of Physics: Condensed Matter}\ }\textbf
  {\bibinfo {volume} {19}},\ \bibinfo {pages} {335219} (\bibinfo {year}
  {2007})}\BibitemShut {NoStop}%
\bibitem [{\citenamefont {Van~Degrift}(1975)}]{TDOReference}%
  \BibitemOpen
  \bibfield  {author} {\bibinfo {author} {\bibfnamefont {C.~T.}\ \bibnamefont
  {Van~Degrift}},\ }\href {https://doi.org/10.1063/1.1134272} {\bibfield
  {journal} {\bibinfo  {journal} {Review of Scientific Instruments}\ }\textbf
  {\bibinfo {volume} {46}},\ \bibinfo {pages} {599} (\bibinfo {year}
  {1975})}\BibitemShut {NoStop}%
\bibitem [{\citenamefont {Shi}\ \emph {et~al.}(2019)\citenamefont {Shi},
  \citenamefont {Steinhardt}, \citenamefont {Graf}, \citenamefont {Corboz},
  \citenamefont {Weickert}, \citenamefont {Harrison}, \citenamefont {Jaime},
  \citenamefont {Marjerrison}, \citenamefont {Dabkowska}, \citenamefont
  {Mila},\ and\ \citenamefont {Haravifard}}]{Shi2019}%
  \BibitemOpen
  \bibfield  {author} {\bibinfo {author} {\bibfnamefont {Z.}~\bibnamefont
  {Shi}}, \bibinfo {author} {\bibfnamefont {W.}~\bibnamefont {Steinhardt}},
  \bibinfo {author} {\bibfnamefont {D.}~\bibnamefont {Graf}}, \bibinfo {author}
  {\bibfnamefont {P.}~\bibnamefont {Corboz}}, \bibinfo {author} {\bibfnamefont
  {F.}~\bibnamefont {Weickert}}, \bibinfo {author} {\bibfnamefont
  {N.}~\bibnamefont {Harrison}}, \bibinfo {author} {\bibfnamefont
  {M.}~\bibnamefont {Jaime}}, \bibinfo {author} {\bibfnamefont
  {C.}~\bibnamefont {Marjerrison}}, \bibinfo {author} {\bibfnamefont {H.~A.}\
  \bibnamefont {Dabkowska}}, \bibinfo {author} {\bibfnamefont {F.}~\bibnamefont
  {Mila}},\ and\ \bibinfo {author} {\bibfnamefont {S.}~\bibnamefont
  {Haravifard}},\ }\href
  {https://doi.org/https://doi.org/10.1038/s41467-019-10410-x} {\bibfield
  {journal} {\bibinfo  {journal} {Nature Communications}\ }\textbf {\bibinfo
  {volume} {10}},\ \bibinfo {pages} {2439} (\bibinfo {year}
  {2019})}\BibitemShut {NoStop}%
\bibitem [{\citenamefont {Azuah}\ \emph {et~al.}(2009)\citenamefont {Azuah},
  \citenamefont {Kneller}, \citenamefont {Qiu}, \citenamefont
  {Tregenna-Piggott}, \citenamefont {Brown}, \citenamefont {Copley},\ and\
  \citenamefont {Dimeo}}]{DAVEMSLICE}%
  \BibitemOpen
  \bibfield  {author} {\bibinfo {author} {\bibfnamefont {R.~T.}\ \bibnamefont
  {Azuah}}, \bibinfo {author} {\bibfnamefont {L.~R.}\ \bibnamefont {Kneller}},
  \bibinfo {author} {\bibfnamefont {Y.}~\bibnamefont {Qiu}}, \bibinfo {author}
  {\bibfnamefont {P.~L.~W.}\ \bibnamefont {Tregenna-Piggott}}, \bibinfo
  {author} {\bibfnamefont {C.~M.}\ \bibnamefont {Brown}}, \bibinfo {author}
  {\bibfnamefont {J.~R.~D.}\ \bibnamefont {Copley}},\ and\ \bibinfo {author}
  {\bibfnamefont {R.~M.}\ \bibnamefont {Dimeo}},\ }\href
  {https://doi.org/10.6028/jres.114.025} {\bibfield  {journal} {\bibinfo
  {journal} {Journal of Research of the National Institute of Standards and
  Technology}\ }\textbf {\bibinfo {volume} {114}},\ \bibinfo {pages} {341}
  (\bibinfo {year} {2009})}\BibitemShut {NoStop}%
\bibitem [{\citenamefont {Arnold}\ \emph {et~al.}(2014)\citenamefont {Arnold},
  \citenamefont {Bilheux}, \citenamefont {Borreguero}, \citenamefont {Buts},
  \citenamefont {Campbell}, \citenamefont {Chapon}, \citenamefont {Doucet},
  \citenamefont {Draper}, \citenamefont {Leal}, \citenamefont {Gigg},
  \citenamefont {Lynch}, \citenamefont {Markvardsen}, \citenamefont
  {Mikkelson}, \citenamefont {Mikkelson}, \citenamefont {Miller}, \citenamefont
  {Palmen}, \citenamefont {Parker}, \citenamefont {Passos}, \citenamefont
  {Perring}, \citenamefont {Peterson}, \citenamefont {Ren}, \citenamefont
  {Reuter}, \citenamefont {Savici}, \citenamefont {Taylor}, \citenamefont
  {Taylor}, \citenamefont {Tolchenov}, \citenamefont {Zhou},\ and\
  \citenamefont {Zikovsky}}]{Mantid}%
  \BibitemOpen
  \bibfield  {author} {\bibinfo {author} {\bibfnamefont {O.}~\bibnamefont
  {Arnold}}, \bibinfo {author} {\bibfnamefont {J.}~\bibnamefont {Bilheux}},
  \bibinfo {author} {\bibfnamefont {J.}~\bibnamefont {Borreguero}}, \bibinfo
  {author} {\bibfnamefont {A.}~\bibnamefont {Buts}}, \bibinfo {author}
  {\bibfnamefont {S.}~\bibnamefont {Campbell}}, \bibinfo {author}
  {\bibfnamefont {L.}~\bibnamefont {Chapon}}, \bibinfo {author} {\bibfnamefont
  {M.}~\bibnamefont {Doucet}}, \bibinfo {author} {\bibfnamefont
  {N.}~\bibnamefont {Draper}}, \bibinfo {author} {\bibfnamefont {R.~F.}\
  \bibnamefont {Leal}}, \bibinfo {author} {\bibfnamefont {M.}~\bibnamefont
  {Gigg}}, \bibinfo {author} {\bibfnamefont {V.}~\bibnamefont {Lynch}},
  \bibinfo {author} {\bibfnamefont {A.}~\bibnamefont {Markvardsen}}, \bibinfo
  {author} {\bibfnamefont {D.}~\bibnamefont {Mikkelson}}, \bibinfo {author}
  {\bibfnamefont {R.}~\bibnamefont {Mikkelson}}, \bibinfo {author}
  {\bibfnamefont {R.}~\bibnamefont {Miller}}, \bibinfo {author} {\bibfnamefont
  {K.}~\bibnamefont {Palmen}}, \bibinfo {author} {\bibfnamefont
  {P.}~\bibnamefont {Parker}}, \bibinfo {author} {\bibfnamefont
  {G.}~\bibnamefont {Passos}}, \bibinfo {author} {\bibfnamefont
  {T.}~\bibnamefont {Perring}}, \bibinfo {author} {\bibfnamefont
  {P.}~\bibnamefont {Peterson}}, \bibinfo {author} {\bibfnamefont
  {S.}~\bibnamefont {Ren}}, \bibinfo {author} {\bibfnamefont {M.}~\bibnamefont
  {Reuter}}, \bibinfo {author} {\bibfnamefont {A.}~\bibnamefont {Savici}},
  \bibinfo {author} {\bibfnamefont {J.}~\bibnamefont {Taylor}}, \bibinfo
  {author} {\bibfnamefont {R.}~\bibnamefont {Taylor}}, \bibinfo {author}
  {\bibfnamefont {R.}~\bibnamefont {Tolchenov}}, \bibinfo {author}
  {\bibfnamefont {W.}~\bibnamefont {Zhou}},\ and\ \bibinfo {author}
  {\bibfnamefont {J.}~\bibnamefont {Zikovsky}},\ }\href
  {https://doi.org/10.1016/j.nima.2014.07.029} {\bibfield  {journal} {\bibinfo
  {journal} {Nuclear Instruments and Methods in Physics Research Section A:
  Accelerators, Spectrometers, Detectors and Associated Equipment}\ }\textbf
  {\bibinfo {volume} {764}},\ \bibinfo {pages} {156} (\bibinfo {year}
  {2014})}\BibitemShut {NoStop}%
\bibitem [{\citenamefont {Molavian}\ \emph {et~al.}(2007)\citenamefont
  {Molavian}, \citenamefont {Gingras},\ and\ \citenamefont
  {Canals}}]{Molavian}%
  \BibitemOpen
  \bibfield  {author} {\bibinfo {author} {\bibfnamefont {H.~R.}\ \bibnamefont
  {Molavian}}, \bibinfo {author} {\bibfnamefont {M.~J.~P.}\ \bibnamefont
  {Gingras}},\ and\ \bibinfo {author} {\bibfnamefont {B.}~\bibnamefont
  {Canals}},\ }\href {https://doi.org/10.1103/PhysRevLett.98.157204} {\bibfield
   {journal} {\bibinfo  {journal} {Phys. Rev. Lett.}\ }\textbf {\bibinfo
  {volume} {98}},\ \bibinfo {pages} {157204} (\bibinfo {year}
  {2007})}\BibitemShut {NoStop}%
\bibitem [{\citenamefont {Aroyo}(2016)}]{inta}%
  \BibitemOpen
  \bibinfo {editor} {\bibfnamefont {M.~I.}\ \bibnamefont {Aroyo}},\ ed.,\ \href
  {https://doi.org/https://doi.org/10.1107/97809553602060000114} {\emph
  {\bibinfo {title} {International Tables for Crystallography. Vol. A, Space
  Group Symmetry}}},\ \bibinfo {edition} {2nd}\ ed.\ (\bibinfo {year}
  {2016})\BibitemShut {NoStop}%
\bibitem [{\citenamefont {Ross}\ \emph {et~al.}(2011)\citenamefont {Ross},
  \citenamefont {Savary}, \citenamefont {Gaulin},\ and\ \citenamefont
  {Balents}}]{RossPRX2011}%
  \BibitemOpen
  \bibfield  {author} {\bibinfo {author} {\bibfnamefont {K.~A.}\ \bibnamefont
  {Ross}}, \bibinfo {author} {\bibfnamefont {L.}~\bibnamefont {Savary}},
  \bibinfo {author} {\bibfnamefont {B.~D.}\ \bibnamefont {Gaulin}},\ and\
  \bibinfo {author} {\bibfnamefont {L.}~\bibnamefont {Balents}},\ }\href
  {https://doi.org/10.1103/PhysRevX.1.021002} {\bibfield  {journal} {\bibinfo
  {journal} {Phys. Rev. X}\ }\textbf {\bibinfo {volume} {1}},\ \bibinfo {pages}
  {021002} (\bibinfo {year} {2011})}\BibitemShut {NoStop}%
\bibitem [{\citenamefont {Wilson}(1992)}]{intc}%
  \BibitemOpen
  \bibinfo {editor} {\bibfnamefont {A.~J.~C.}\ \bibnamefont {Wilson}},\ ed.,\
  \href@noop {} {\emph {\bibinfo {title} {International Tables for
  Crystallography. Vol. C, Mathematical, physical and chemical tables}}}\
  (\bibinfo  {publisher} {Kluwer Academic Publishers},\ \bibinfo {address}
  {Dordrecht/Boston/London},\ \bibinfo {year} {1992})\BibitemShut {NoStop}%
\bibitem [{\citenamefont {Haravifard}\ \emph {et~al.}(2016)\citenamefont
  {Haravifard}, \citenamefont {Graf}, \citenamefont {Feiguin}, \citenamefont
  {Batista}, \citenamefont {Lang}, \citenamefont {Silevitch}, \citenamefont
  {Srajer}, \citenamefont {Gaulin}, \citenamefont {Dabkowska},\ and\
  \citenamefont {Rosenbaum}}]{Haravifard2016}%
  \BibitemOpen
  \bibfield  {author} {\bibinfo {author} {\bibfnamefont {S.}~\bibnamefont
  {Haravifard}}, \bibinfo {author} {\bibfnamefont {D.}~\bibnamefont {Graf}},
  \bibinfo {author} {\bibfnamefont {A.~E.}\ \bibnamefont {Feiguin}}, \bibinfo
  {author} {\bibfnamefont {C.~D.}\ \bibnamefont {Batista}}, \bibinfo {author}
  {\bibfnamefont {J.~C.}\ \bibnamefont {Lang}}, \bibinfo {author}
  {\bibfnamefont {D.~M.}\ \bibnamefont {Silevitch}}, \bibinfo {author}
  {\bibfnamefont {G.}~\bibnamefont {Srajer}}, \bibinfo {author} {\bibfnamefont
  {B.~D.}\ \bibnamefont {Gaulin}}, \bibinfo {author} {\bibfnamefont {H.~A.}\
  \bibnamefont {Dabkowska}},\ and\ \bibinfo {author} {\bibfnamefont {T.~F.}\
  \bibnamefont {Rosenbaum}},\ }\href
  {https://doi.org/https://doi.org/10.1038/ncomms11956} {\bibfield  {journal}
  {\bibinfo  {journal} {Nature Communications}\ }\textbf {\bibinfo {volume}
  {7}},\ \bibinfo {pages} {11956} (\bibinfo {year} {2016})}\BibitemShut
  {NoStop}%
\bibitem [{\citenamefont {Steinhardt}\ \emph {et~al.}(2021)\citenamefont
  {Steinhardt}, \citenamefont {Shi}, \citenamefont {Samarakoon}, \citenamefont
  {Dissanayake}, \citenamefont {Graf}, \citenamefont {Liu}, \citenamefont
  {Zhu}, \citenamefont {Marjerrison}, \citenamefont {Batista},\ and\
  \citenamefont {Haravifard}}]{Steinhardt2021}%
  \BibitemOpen
  \bibfield  {author} {\bibinfo {author} {\bibfnamefont {W.}~\bibnamefont
  {Steinhardt}}, \bibinfo {author} {\bibfnamefont {Z.}~\bibnamefont {Shi}},
  \bibinfo {author} {\bibfnamefont {A.}~\bibnamefont {Samarakoon}}, \bibinfo
  {author} {\bibfnamefont {S.}~\bibnamefont {Dissanayake}}, \bibinfo {author}
  {\bibfnamefont {D.}~\bibnamefont {Graf}}, \bibinfo {author} {\bibfnamefont
  {Y.}~\bibnamefont {Liu}}, \bibinfo {author} {\bibfnamefont {W.}~\bibnamefont
  {Zhu}}, \bibinfo {author} {\bibfnamefont {C.}~\bibnamefont {Marjerrison}},
  \bibinfo {author} {\bibfnamefont {C.~D.}\ \bibnamefont {Batista}},\ and\
  \bibinfo {author} {\bibfnamefont {S.}~\bibnamefont {Haravifard}},\ }\href
  {https://doi.org/10.1103/PhysRevResearch.3.033050} {\bibfield  {journal}
  {\bibinfo  {journal} {Phys. Rev. Research}\ }\textbf {\bibinfo {volume}
  {3}},\ \bibinfo {pages} {033050} (\bibinfo {year} {2021})}\BibitemShut
  {NoStop}%
\bibitem [{Note1()}]{Note1}%
  \BibitemOpen
  \bibinfo {note} {Depending on the magnitude of the field sweep rate relative
  to the energy splitting at the avoid crossing, non-adiabatic transitions
  between the ground and first excited states are possible and their
  probability can be estimated using the usual Landau-Zener formula.
  Significant non-adiabatic probability implies hysteresis in the state of the
  system (and thus magnetization) as the field is swept up or
  down.}\BibitemShut {Stop}%
\bibitem [{Note2()}]{Note2}%
  \BibitemOpen
  \bibinfo {note} {The authors of Ref.~\cite {Park2016} attribute the
  hysteresis observed in their data to a non-equilibrium effect due to the
  finite rate of change in the magnetic field strength. However, estimates of
  the probability of non-adiabatic transitions (using the standard Laudau-Zener
  formula~\cite {landau1987quantum}) for any reasonable splitting size
  [$O({\protect \rm meV})$] and field sweep rate [$O({\protect \rm mT/min}]$]
  are negligible.}\BibitemShut {Stop}%
\bibitem [{\citenamefont {Takeshi}\ and\ \citenamefont
  {Billinge}(2012)}]{TAKESHI201255}%
  \BibitemOpen
  \bibfield  {author} {\bibinfo {author} {\bibfnamefont {E.}~\bibnamefont
  {Takeshi}}\ and\ \bibinfo {author} {\bibfnamefont {S.~J.}\ \bibnamefont
  {Billinge}},\ }in\ \href
  {https://doi.org/https://doi.org/10.1016/B978-0-08-097133-9.00003-4} {\emph
  {\bibinfo {booktitle} {Underneath the Bragg Peaks}}},\ \bibinfo {series}
  {Pergamon Materials Series}, Vol.~\bibinfo {volume} {16},\ \bibinfo {editor}
  {edited by\ \bibinfo {editor} {\bibfnamefont {T.}~\bibnamefont {Egami}}\ and\
  \bibinfo {editor} {\bibfnamefont {S.~J.}\ \bibnamefont {Billinge}}}\
  (\bibinfo  {publisher} {Pergamon},\ \bibinfo {year} {2012})\ pp.\ \bibinfo
  {pages} {55--111}\BibitemShut {NoStop}%
\bibitem [{Note3()}]{Note3}%
  \BibitemOpen
  \bibinfo {note} {We note that this additional symmetry arises due to the
  symmetrization of the unpolarized intensity and the integration of over
  energy transfer that reduces the structure factor to static moment-moment
  correlation function.}\BibitemShut {Stop}%
\bibitem [{\citenamefont {Mirebeau}\ \emph {et~al.}(2002)\citenamefont
  {Mirebeau}, \citenamefont {Goncharenko}, \citenamefont {Cadavez-Pares},
  \citenamefont {Bramwell}, \citenamefont {Gingras},\ and\ \citenamefont
  {Gardner}}]{MirebeauNature}%
  \BibitemOpen
  \bibfield  {author} {\bibinfo {author} {\bibfnamefont {I.}~\bibnamefont
  {Mirebeau}}, \bibinfo {author} {\bibfnamefont {I.~N.}\ \bibnamefont
  {Goncharenko}}, \bibinfo {author} {\bibfnamefont {P.}~\bibnamefont
  {Cadavez-Pares}}, \bibinfo {author} {\bibfnamefont {S.~T.}\ \bibnamefont
  {Bramwell}}, \bibinfo {author} {\bibfnamefont {M.~J.~P.}\ \bibnamefont
  {Gingras}},\ and\ \bibinfo {author} {\bibfnamefont {J.~S.}\ \bibnamefont
  {Gardner}},\ }\href {https://doi.org/10.1038/nature01157} {\bibfield
  {journal} {\bibinfo  {journal} {Nature}\ }\textbf {\bibinfo {volume} {420}},\
  \bibinfo {pages} {54} (\bibinfo {year} {2002})}\BibitemShut {NoStop}%
\bibitem [{\citenamefont {Mirebeau}\ \emph {et~al.}(2004)\citenamefont
  {Mirebeau}, \citenamefont {Goncharenko}, \citenamefont {Dhalenne},\ and\
  \citenamefont {Revcolevschi}}]{MirebeauPRL}%
  \BibitemOpen
  \bibfield  {author} {\bibinfo {author} {\bibfnamefont {I.}~\bibnamefont
  {Mirebeau}}, \bibinfo {author} {\bibfnamefont {I.~N.}\ \bibnamefont
  {Goncharenko}}, \bibinfo {author} {\bibfnamefont {G.}~\bibnamefont
  {Dhalenne}},\ and\ \bibinfo {author} {\bibfnamefont {A.}~\bibnamefont
  {Revcolevschi}},\ }\href {https://doi.org/10.1103/PhysRevLett.93.187204}
  {\bibfield  {journal} {\bibinfo  {journal} {Phys. Rev. Lett.}\ }\textbf
  {\bibinfo {volume} {93}},\ \bibinfo {pages} {187204} (\bibinfo {year}
  {2004})}\BibitemShut {NoStop}%
\bibitem [{\citenamefont {Landau}\ and\ \citenamefont
  {Lifshitz}(1987)}]{landau1987quantum}%
  \BibitemOpen
  \bibfield  {author} {\bibinfo {author} {\bibfnamefont {L.~D.}\ \bibnamefont
  {Landau}}\ and\ \bibinfo {author} {\bibfnamefont {E.~M.}\ \bibnamefont
  {Lifshitz}},\ }\href@noop {} {\emph {\bibinfo {title} {Quantum Mechanics,
  Course of Theoretical Physics, Vol.3}}}\ (\bibinfo {year} {1987})\ p.\
  \bibinfo {pages} {167}\BibitemShut {NoStop}%
\end{thebibliography}%

\end{document}